\documentclass[aps,pre,reprint,superscriptaddress,longbibliography]{revtex4-2}
\usepackage{amsmath,amssymb}
\usepackage{graphicx}
\usepackage{hyperref}
\usepackage{xcolor}
\usepackage{svg}
\usepackage{soul, color}
\usepackage[normalem]{ulem}

\newcommand{\figwidth}{\linewidth} 
\newcommand{\G}[1]{\textcolor{black}{#1}}
\newcommand{\R}[1]{\textcolor{black}{#1}}
\newcommand{\B}[1]{\textcolor{black}{#1}}

\begin{document}

\title{Phase Transitions in Economic Inequality:\\
Taxation and Extremal Replacement Dynamics}

\author{Lautaro Giordano}
\thanks{Corresponding author: lautaro.giordano@ib.edu.ar}
\affiliation{Statistical and Interdisciplinary Physics Group, Centro Atómico Bariloche (CNEA) and CONICET. R8402AGP Bariloche, Argentina.}
\affiliation{Instituto Balseiro, Universidad Nacional de Cuyo. R8402AGP Bariloche, Argentina.}

\author{Sebastian Gonçalves}
\affiliation{Instituto de Física, Universidade Federal do Rio Grande do Sul, 91501-970 Porto Alegre RS, Brazil.}

\author{José Roberto Iglesias}
\affiliation{Instituto de Física, Universidade Federal do Rio Grande do Sul, 91501-970 Porto Alegre RS, Brazil.}
\affiliation{Instituto Nacional de Ciência e Tecnologia de Sistemas Complexos, CBPF, Rio de Janeiro, Brazil.}

\author{María Fabiana Laguna}
\affiliation{Statistical and Interdisciplinary Physics Group, Centro Atómico Bariloche (CNEA) and CONICET. R8402AGP Bariloche, Argentina.}

\begin{abstract}
We present a minimal agent-based model of interacting agents characterized by their wealth to study taxation and inequality in a non-conservative economy. Wealth evolves through an extremal stochastic replacement process in which the poorest agent has its wealth replaced by a new random value, financed through a collective taxation mechanism. We explore taxation regimes ranging from regressive to progressive schemes and tune the overall redistribution strength.
Under regressive taxation, the system self-organizes into two distinct stationary phases when changing the total tax collected: a non-ergodic, high-inequality regime characterized by wealth condensation in a subset of agents that permanently escape replacement, and a more homogeneous ergodic phase in which all agents participate in the dynamics. Increasing taxes drives an abrupt transition between these phases. The transition is discontinuous and exhibits hysteresis and bistability, consistently detected through the Gini index, the Top $1\%$ wealth share, the entropy, and the Binder cumulant.
In contrast, neutral and progressive taxation suppress persistent wealth concentration, preventing the emergence of strongly unequal states and eliminating hysteretic behavior. We complement the simulations with a parameter-free mean-field theory of the wealth dynamics that reproduces the stationary distributions and explains the transition by locating the two spinodals that bound the bistable region.
These results show that minimal stochastic redistribution mechanisms alone can produce discontinuous transitions, metastability, and non-ergodicity, demonstrating that taxation structure can determine the emergence and stability of macroscopic inequality.
\end{abstract}

\maketitle

\section{Introduction} 
\label{sec:intro}

Economic inequality has long been studied not only in economics and sociology, but also in political philosophy. 
\B{According to Rawls's difference principle \cite{RAWLS1971}, inequalities are justified only when they improve the welfare of the least advantaged members of society. Although originally formulated as a normative principle, this perspective naturally motivates the study of dynamical mechanisms that act on the least advantaged agents and their consequences for the evolution of wealth inequality.}

\B{
Empirical evidence accumulated over recent decades has documented the persistence and amplification of wealth inequality in capitalist economies \cite{Piketty2014}, 
motivating the development of  minimal microscopic models capable of reproducing highly unequal stationary states.
}

\B{
In parallel with these normative and empirical approaches, the statistical physics of wealth and income distributions has developed a broad family of agent-based and kinetic exchange models aimed at understanding the emergence of inequality from microscopic interaction rules \cite{Pareto1897,YakovenkoRosser2009,Chakrabarti2013, MantegnaStanley2000, Lux2008}. Early stochastic exchange approaches include the surplus theory of social stratification introduced by Angle \cite{Angle1986,Angle1992}. Subsequent developments gave rise to several major classes of econophysical models of wealth dynamics: conservative pairwise exchange models \cite{Dragulescu2000}; saving-propensity models \cite{Chakraborti2000,Chatterjee2004}; multiplicative-growth and network-based approaches \cite{BouchaudMezard2000,IGLESIAS2004}; non-conservative or open-economy formulations \cite{Slanina2004}; extremal-dynamics models inspired by self-organized criticality \cite{PIANEGONDA2003}; and taxation or redistribution models \cite{Boghosian2017,Liu2021PhysRevE}. Together, these approaches have revealed a rich variety of stationary states, inequality regimes, and phase-transition-like phenomena \cite{YakovenkoRosser2009,Chakrabarti2013,Lux2008}.
Our model combines ingredients from the last two classes: it is based on extremal replacement dynamics, but incorporates collective taxation and redistribution mechanisms in a non-conservative setting.}

From a modeling perspective, this focus on the least advantaged agent has been naturally incorporated in agent-based models inspired by extremal dynamics. 
The model we present here draws inspiration from the work by Pianegonda et al.~\cite{PIANEGONDA2003}, itself motivated by the extremal dynamics model of extinction of species by Bak and Sneppen~\cite{BakSneppen1993}. In the original framework, each agent is assigned an initial wealth value drawn uniformly from the interval [0,1), with the focus on the poorest agent, namely the one holding the minimum wealth. This agent attempts to improve its economic standing through a stochastic action: its wealth is reassigned by drawing a new value uniformly from [0,1). This update can also be interpreted as the poorest agent defaulting, being removed from the economic system, and replaced by a new agent with random wealth. 

Unlike the Bak--Sneppen model, the Pianegonda et al. model is conservative, hence its name: the Conservative Exchanges Market Model (CEMM). Within the CEMM, the change in wealth of the poorest agent is exactly compensated by redistributing the opposite amount in equal fractions among a subset of other agents. This group may range from just one or two agents to the entire population. Therefore, any improvement in the economic condition of the poorest agent necessarily comes at the expense of others, who collectively lose the same total amount of wealth. This replacement mechanism captures individual economic mobility under uncertainty while preserving computational simplicity. 

Variants of this model, differing in the number of agents involved in exchanges, or in the network structure \cite{IGLESIAS2003}, display consistent features: nearly all agents reside above a poverty threshold (whose exact value depends on model details), with those below it most likely being the minimum-wealth agents, whose state will change in subsequent simulation steps. Additionally, the wealth distribution rises from zero at the poverty line, peaks sharply, and then decays exponentially, with an exponent proportional to the square of the wealth. 

In the present work, we invert the logic of the inspirational model. Rather than modifying the poorest agent’s wealth and redistributing the difference among neighbors, we impose a tax on all agents in this hypothetical society. 
\R{
By assigning a privileged dynamical role to the least-wealth agent, the model is loosely inspired by Rawlsian ideas that place particular emphasis on the condition of the least advantaged members of society \cite{RAWLS1971}. This connection should be understood only at a conceptual level, since the model is not intended as a realistic representation of any specific redistribution policy.}
The collected revenue is not distributed for the benefit of a subset of agents, as commonly implemented in other models \citep{BISI2009, DINIZ2012, BOUCHAUD2015, DIAS2024}, but is instead coupled to an extremal replacement dynamics in which the poorest agent is removed from the system and replaced by a new agent with randomly assigned wealth.
\R{In this sense, the taxation mechanism acts as a stochastic resource injection associated with the replacement process rather than a literal welfare transfer.}
Furthermore, the agent’s previous wealth is discarded upon replacement, making the dynamics non-conservative: except when the poorest agent is initially in debt, total wealth decreases after each step. As shown below, this latter case is rare, so the dominant effect is a systematic loss of total wealth, which we renormalize to avoid numerical underflow. These seemingly small modifications \B{(pooling tax revenues, coupling them to the replacement of the least-wealth agent, and abandoning wealth conservation) lead to qualitatively new emergent behavior.}

\B{
In standard exchange models with pairwise interactions, including the random exchange model of Dragulescu and Yakovenko \cite{Dragulescu2000} and the saving-propensity model introduced by Chakraborti and Chakrabarti \cite{Chakraborti2000}, wealth is redistributed through local transactions that conserve total resources and typically lead to stationary distributions whose inequality varies smoothly with model parameters. These models, later popularized in the context of the ``Yard-Sale'' dynamics discussed by Hayes \cite{Hayes2002}, became central paradigms in the econophysics literature on wealth inequality.
}

Extensions incorporating saving propensity heterogeneity \cite{Chatterjee2004}, multiplicative growth \cite{BouchaudMezard2000}, risk strategies~\cite{Giordano2025}, or network structure \cite{IGLESIAS2004} likewise preserve this continuous dependence. 

\B{
More recent formulations incorporating taxation, redistribution, stochastic growth, or wealth-attained advantages 
\cite{Boghosian2017, LI2019,Liu2021PhysRevE,Nener2021PhysRevE,Calvelli2003,Villafane2025,VazquezVonBibow2025PhysRevE} have introduced additional economic realism while retaining the general feature that macroscopic inequality indicators evolve continuously as control parameters are tuned.
}

\R{
Phase transitions, wealth-condensation phenomena, and broad wealth distributions have been reported in asset-exchange models based on minimum-exchange or Yard-Sale dynamics \cite{Sinha2003,Liu2021PhysRevE}. In those cases, the transitions are typically associated with multiplicative exchange mechanisms, exchange efficiency, or redistribution rates. Here we show that qualitatively similar macroscopic phenomena, namely abrupt transitions, hysteresis, and bistability, can arise from a different mechanism: the interplay between taxation and extremal replacement dynamics in a non-conservative setting. In particular, we show that the structure of taxation alone can trigger macroscopic reorganizations of wealth exhibiting features commonly associated with discontinuous transitions in physical systems.
}

The outline of the article is as follows: In the next section, we describe the model in detail and define the indicators used to characterize inequality; we then present the numerical results, discussing the emergence of distinct dynamical regimes and phase transitions under different taxation schemes. \G{We complement these results with a mean-field theory that reproduces the stationary distributions and explains the transition by locating the two spinodals that bound the coexistence region.} Finally, we conclude with a summary of the results and a discussion of their broader implications.

\section{Model} 
\label{sec:model}

The system is composed of $N$ agents characterized by their wealth $w$, initially drawn from a uniform distribution in the range $[0,1)$, and subsequently rescaled so that the total wealth satisfies $W = \sum_{i=1}^{N}w_i = 1$. The system evolves in discrete time through two sequential processes: a \textit{tax collection step} and a \textit{wealth replacement step}.
In broad terms, a fraction of the system's wealth is extracted through taxation. Subsequently, an agent is selected according to an extremal criterion, and its wealth is reassigned to a new value drawn from a \B{probability }distribution \B{whose parameters depend on the tax revenue collected}. 
This mechanism \B{provides a minimal representation of redistribution in which the resources collected through taxation are directed toward the least wealthy agent.}

\B{Consistent with many agent-based and Monte Carlo models, time is measured in terms of elementary update events of the stochastic dynamics and is not associated with any specific empirical timescale.}

\subsection{Tax collection step}

At each timestep, a total tax 
\begin{equation}
T=\frac{\gamma\langle w\rangle}{2}
\end{equation}
is collected from all agents with positive wealth, where $\gamma>0$ is called the tax strength parameter and 
\begin{equation}
    \langle w\rangle = \frac{1}{N} \sum_{i=1}^{N}w_i = \frac{W}{N}
\end{equation}
is the mean wealth of the system, proportional to the total wealth $W$. The magnitude of the tax depends on the strength parameter and scales with the system's wealth. The tax paid by agent $i$ is given by
\begin{equation}
T_i = 
\begin{cases} 
\frac{w_i^\alpha}{\sum\limits_{w_j>0} w_j^\alpha}\,T & \text{if } w_i > 0\\
0 & \text{if } w_i \leq 0
\end{cases}
\end{equation}
The parameter $\alpha \geq 0$ is called the tax progressivity, and it determines the weight assigned to each taxpayer. When $\alpha < 1$ the tax is regressive, i.e., poorer agents contribute proportionally more to the total tax (in the special case of $\alpha=0$ the tax is flat and all positive-wealth agents pay the same amount). For $\alpha>1$ the tax is progressive, as agents contribute proportionally more to the tax as their wealth increases. Agents with negative wealth do not pay taxes. In the explored parameter range, the stationary fraction of agents with $w\leq0$ is negligible (typically zero after transients), so ${\sum_{w_j>0} w_j^\alpha}\approx{\sum_{j} w_j^\alpha}$.

\subsection{Wealth replacement step}

Following the principles of extremal dynamics, the poorest agent in the system, with wealth $w_p(t)$, is selected. Its wealth will be replaced by a new value drawn from a uniform distribution in the interval $[0, 2T)$,
\begin{equation}
    \label{eq:poorestReplacement}
    w_p(t+1) = \mathrm{U}[0, 2T) = \mathrm{U}[0,\gamma\langle w\rangle).
\end{equation}

The parameter $\gamma$ controls the strength of redistribution. For instance, $\gamma = 2$ implies that the poorest agent’s post-update wealth can take values up to twice the mean wealth of the system. The factor $\langle w \rangle$ provides a dynamic rescaling, as total wealth is not strictly conserved in the model.
The collected tax revenue determines the distribution from which the poorest agent's new wealth is sampled. The new wealth is drawn uniformly from $[0,2T)$, so its expected value equals the total collected tax $T$. However, individual realizations may fall below or above $T$, representing heterogeneous outcomes of redistributive policies ranging from partial dissipation (e.g., urgent consumption or debt repayment) to successful recovery and upward mobility.

\R{Importantly, the previous wealth of the poorest agent is discarded when the new value is assigned. Consequently, the dynamics is non-conservative and constitutes a renewal process coupled to taxation: depending on the realized value of the reset, wealth may be effectively dissipated or generated at each step.}
\B{The change in the total wealth after one update is}
\begin{equation}\label{eq:W(t+1)}
    \Delta W = - [T + (w_p(t)-T_p(t))] + \mathrm{U}[0,2)T\,.
\end{equation}
\B{Averaging over the stochastic replacement, the average change in total wealth depends only on the original wealth of the poorest agent after taxation,}
\begin{equation}
\label{eq:ExpectedTotWChange}
\mathbb{E}[\Delta W] = - w_p + T_p.
\end{equation}
Since the poorest agent’s wealth is not constrained to remain positive, the total wealth may either decrease or increase depending on the realized replacement value and on the taxation parameters. In practice, for most parameter values explored in this study, the total wealth exhibits a decreasing trend over time \B{and therefore does not reach a stationary value. Consequently, the absolute wealth is not the relevant quantity for characterizing the long-term dynamics. Instead, after each timestep we normalize the agents' wealth by} the instantaneous total wealth,
\begin{equation}
    \label{eq:normStep}
    w_i(t^+) \;\rightarrow\; \frac{w_i(t^+)}{W(t^+)}, 
\end{equation}
where $t^+$ indicates the time after the tax collection and wealth replacement steps have been completed. The normalized wealths satisfy $W(t)=1$ for all $t$, after the normalization step. This normalization corresponds to a global rescaling of all agents’ wealth and therefore preserves the relative ordering of agents and the shape of the wealth distribution. Therefore, scale-free observables that depend only on relative shares are invariant under this rescaling, which merely stabilizes numerical values in the long-time dynamics. We verified numerically that the stationary indicators considered in this work are unchanged when the dynamics is performed without the normalization.

To measure the level of inequality, the Gini index is widely used. For a discrete system of agents, it is defined as
\begin{equation}
    G(t) = \frac{1}{2NW(t)} \sum_{i,j=1}^{N}|w_i(t)-w_j(t)|.
\end{equation}
This index ranges from 0 (perfect equality) to 1 (complete concentration of wealth in a single agent). Another measure included in this work is the Redistributive Effect index (RE index),
\begin{equation}
\label{eqRE}
    RE(t) = G_{\textrm{pre}}(t) - G_{\textrm{post}}(t),
\end{equation}
which accounts for the change in the Gini index after the tax is collected. The sign of $RE$ directly indicates whether the tax scheme increases or decreases inequality. While this definition provides a direct and easily interpretable quantification of the instantaneous impact of taxation on inequality, it is formally analogous to the Reynolds--Smolensky index and conceptually related to standard redistributive measures widely used in economics~\cite{ReynoldsSmolensky1977,Lambert2001,Kakwani1977}.

Simulations were run for at least $10^5$ timesteps and continued until stationarity was reached. Stationarity was assessed by comparing the mean Gini index computed over the last $25{,}000$ steps with that obtained over the preceding $25{,}000$ steps. \B{The system was considered stationary when the difference between the two averages was smaller than $0.005$.} As anticipated above, the Gini index is invariant under the normalization step introduced in Eq.~\ref{eq:normStep}, ensuring that normalization does not affect the measured level of inequality nor the identification of stationary regimes.

Unless otherwise stated, results are reported for systems of size $N=1000$ agents. The model behavior is analyzed for several values of the parameters $\alpha$ and $\gamma$, which respectively control the taxation regime and its overall strength. These parameters are treated as control variables and explored over the ranges $\alpha \in [0,2]$ and $\gamma \in [0.1,3]$, allowing us to characterize the different dynamical regimes exhibited by the system.

\section{Results} 
\label{sec:results}

\subsection{Characterization of the model dynamics} 
\label{subsec:characterization}

\begin{figure}[ht]
\centering
\includegraphics[width=\figwidth]{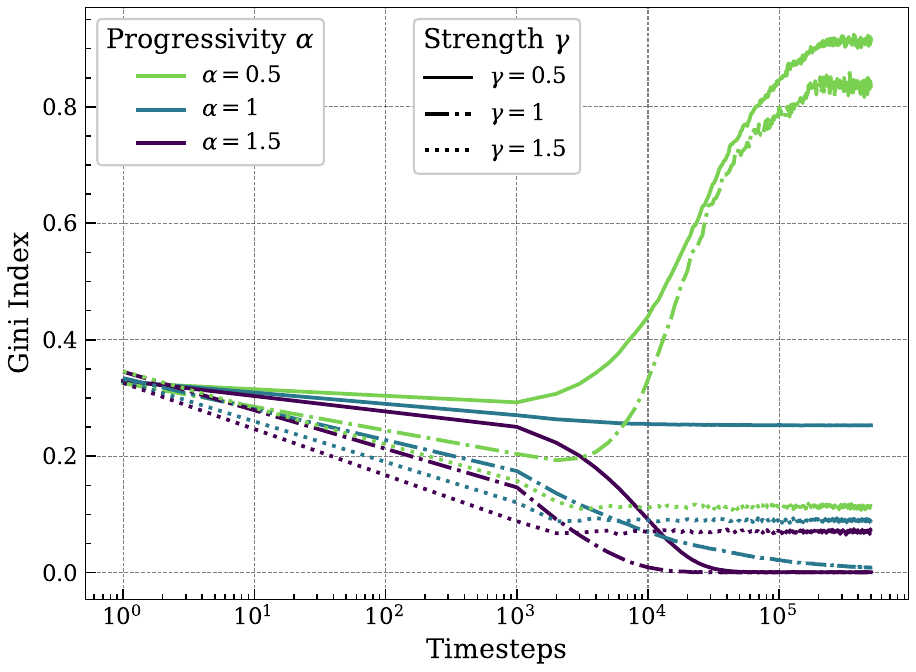}
\caption{\B{Temporal} evolution of the Gini index for individual systems at selected values of $\alpha$ and $\gamma$. Colors indicate the tax progressivity ($\alpha$) value, while line styles correspond to different tax strengths ($\gamma$). Parameter values are specified in the legend.
All systems were run for $5\times10^5$ time steps.
}
\label{fig:giniEvo}
\end{figure}

To explore the dynamical behavior of the system, we first analyze in Fig.~\ref{fig:giniEvo} the temporal evolution of the Gini index for different combinations of the control parameters, recalling that $\gamma$ is related to the new wealth of the poorest agent and the total tax collected, and $\alpha$ sets the structure of such a tax.

In all cases, the system reaches a stationary state, as defined by the Gini-based criterion introduced in the previous section, after a transient whose duration depends on the parameters. The approach to stationarity is typically faster for progressive taxes ($\alpha >1$) and moderate values of $\gamma$, while for regressive taxation ($\alpha < 1$) the relaxation becomes slower due to the emergence of large fluctuations in the wealth distribution. All curves eventually satisfy the stationary condition, yielding asymptotic Gini values that span a wide range across the parameter space.

\begin{figure}[ht]
\centering
\includegraphics[width=\figwidth]
{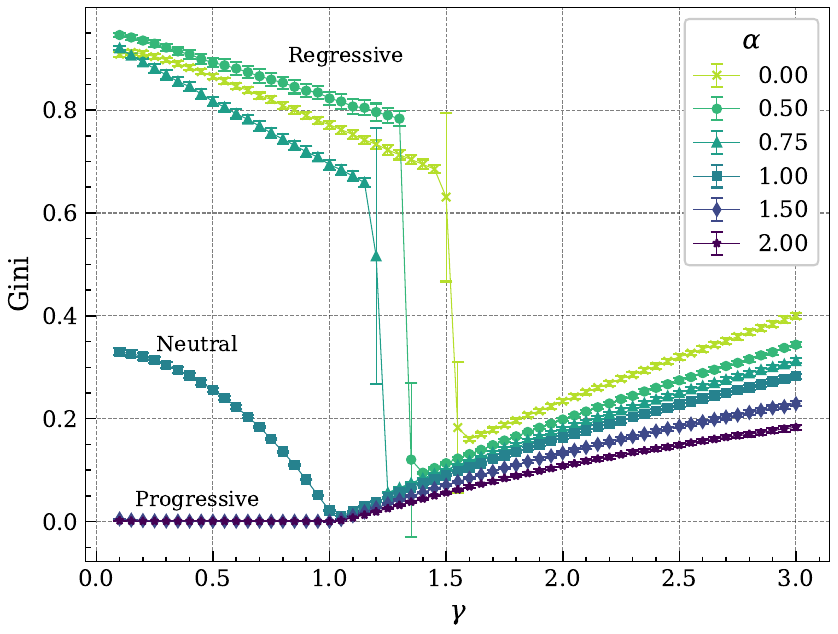}
\caption{Stationary Gini index as a function of $\gamma$ for different values of $\alpha$. Each point is the average over 100 independent realizations after the system reached the stationary state, with the corresponding error bars indicating the standard deviation. Lines are shown as a guide to the eye.}
\label{fig:giniFinalAlphas}
\end{figure}

After the system reaches a steady state, the dependence of inequality on the model parameters can be examined through the Gini index. Figure~\ref{fig:giniFinalAlphas} shows the stationary Gini index as a function of $\gamma$ for different values of $\alpha$. The curves reveal three distinct regimes associated with the taxation regime.

For regressive taxation ($\alpha < 1$), the Gini index exhibits a discontinuous jump from a high-inequality state to a much lower value once $\gamma$ exceeds a parameter-dependent threshold. This abrupt change indicates a transition between distinct stationary regimes, analyzed in detail in Secs.~\ref{subsec:phaseTransition}~and~\ref{subsec:spinodals}. The large error bars observed in this region for $\alpha < 1$ suggest the presence of multiple dynamically accessible stationary states.

For progressive taxation ($\alpha > 1$), the Gini index remains close to zero up to $\gamma \approx 1$. In this regime of progressive taxation and moderate redistribution strength, the system reaches a state of almost perfect equality, in which all agents hold approximately the same wealth. For $\gamma > 1$, the Gini index displays a slow but monotonic increase with $\gamma$, remaining at relatively low inequality levels. 
\R{This behavior reflects the increasing importance of the replacement mechanism: as $\gamma$ grows, the wealth assigned to the poorest agent is drawn from a progressively broader interval, generating larger fluctuations in individual wealth. These fluctuations increase the characteristic width of the stationary wealth distribution and consequently produce a moderate increase in the Gini index, even under progressive taxation. Nevertheless, the evolution remains smooth throughout the explored parameter range and no discontinuous transition is observed.}
\R{The smooth evolution observed for $\alpha \ge 1$ contrasts sharply with the abrupt transition found for $\alpha < 1$, highlighting the ability of progressive taxation to prevent the emergence of the strongly unequal regime observed under regressive taxation.}
We have also verified that the stationary inequality indicators are essentially independent of system size for $N$ ranging from $500$ to $10^4$ (see Appendix \ref{app:finitesize}).

The neutral case ($\alpha = 1$) displays an intermediate behavior between the progressive and regressive regimes. The Gini index decreases with increasing $\gamma$ up to $\gamma \approx 1$, where it reaches a minimum, and then increases smoothly for larger $\gamma$, without exhibiting a discontinuous transition. Interestingly, the minimum occurs close to $\gamma = 1$, where the maximum attainable wealth after replacement becomes comparable to the mean wealth of the system, a feature that will become relevant later in our analysis.

\begin{figure}[ht]
\centering
\includegraphics[width=\figwidth]{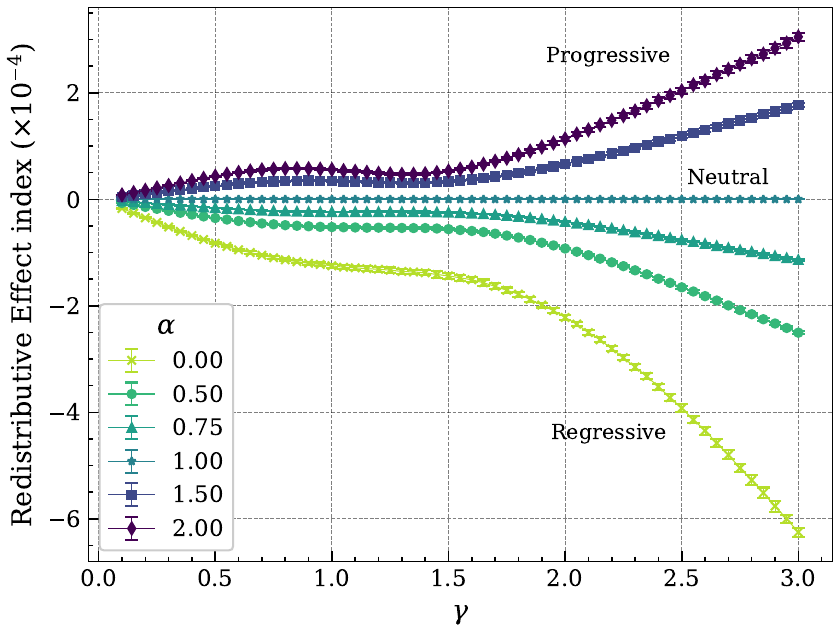}
\caption{Stationary Redistributive Effect index as a function of $\gamma$, for varying $\alpha$. 
 Each point is the average over 100 independent realizations of the RE index, after the system reached the stationary state. Error bars represent the corresponding standard deviation.}
\label{fig:REindex}
\end{figure}

The effectiveness of taxation can be further assessed through the Redistributive Effect index (Eq.~\ref{eqRE}), which measures the change in the Gini index induced by the tax collection step, shown in Fig.~\ref{fig:REindex}. Each data point corresponds to the stationary value of the RE index, averaged over 100 realizations, indicating that the RE index remains persistently positive or negative depending on the tax structure.
This quantity is negative for regressive taxes, indicating that inequality increases after tax collection, null for the neutral case, and positive for progressive taxes.
Therefore, the sign of the RE index provides a direct diagnostic of whether the implemented scheme reduces or amplifies inequality. The existence of a stationary state is not in contradiction with nonzero values of this index, as the dynamics results from a balance between the replacement mechanism and the taxation step, whose competing effects stabilize the Gini index.
Moreover, the magnitude of the RE index increases with $\gamma$ (or decreases in the regressive case), reflecting the growing total amount of wealth extracted through taxation.

\begin{figure}[ht]
\centering
\includegraphics[width=\figwidth]{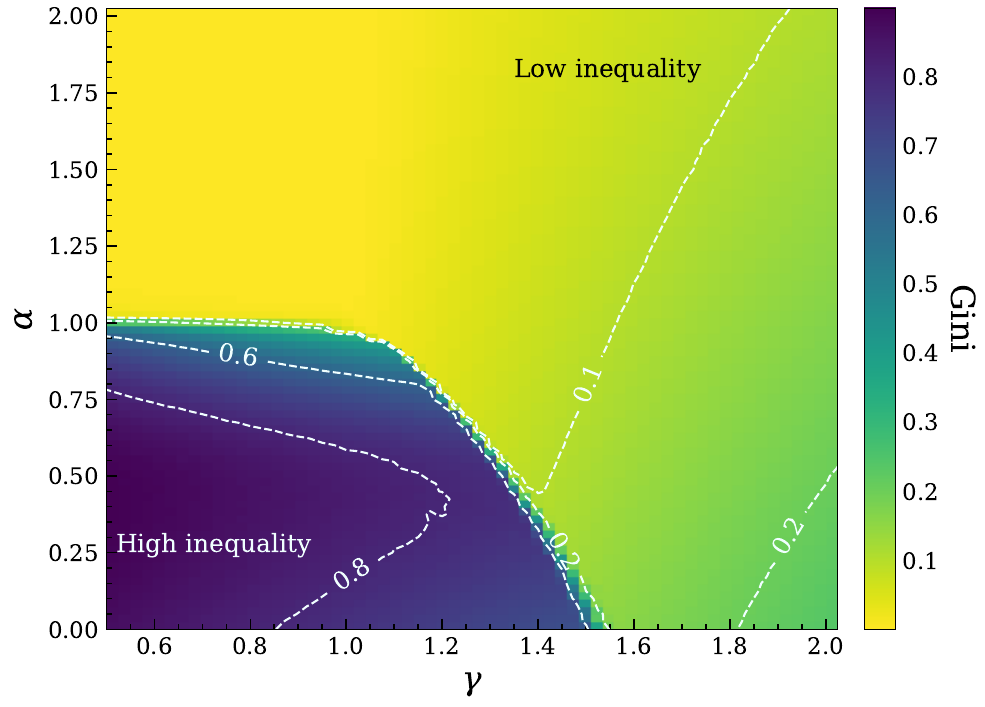}
\caption{
\R{Phase diagram showing the stationary Gini index in the $(\gamma,\alpha)$ plane. Colors represent the stationary Gini value averaged over 100 realizations, while dashed lines denote iso-Gini contours. The high-inequality region occurs at small $\gamma$ for $\alpha<1$. Although the Gini index increases again for large $\gamma$, this smooth increase is qualitatively different from the high-inequality phase.
}}
\label{fig:giniheatmap}
\end{figure}

To further characterize the model results, Fig.~\ref{fig:giniheatmap} shows the phase diagram, mapping the steady-state Gini index in the $(\gamma,\alpha)$ plane.
Two main regions can be identified: a high-inequality phase occurring at small $\gamma$ for $\alpha < 1$, and a low-inequality phase elsewhere.

The transition between these regimes is abrupt when $\gamma$ is varied at fixed $\alpha < 1$, with the Gini index displaying a sudden jump. An analogous behavior is observed when varying $\alpha$ at fixed $\gamma$ within the regressive regime, although this dependence is not shown here for brevity. In both cases, the system exhibits a sharp change between high- and low-inequality stationary states. 

\R{Moreover, for $\alpha>1$ the high-inequality condensed states are effectively suppressed throughout the explored range of $\gamma$. The Gini index does grow with $\gamma$ in this regime (see Fig.~\ref{fig:giniFinalAlphas}), but the rise is smooth and monotonic, with no abrupt jump or change of behavior, reflecting the fluctuation-broadening mechanism discussed above rather than a reappearance of the high-inequality phase. We verified that this trend continues up to $\gamma\sim20$, where the Gini index gradually flattens (not shown here).}

\begin{figure}[ht]
\centering
\includegraphics[width=\figwidth]{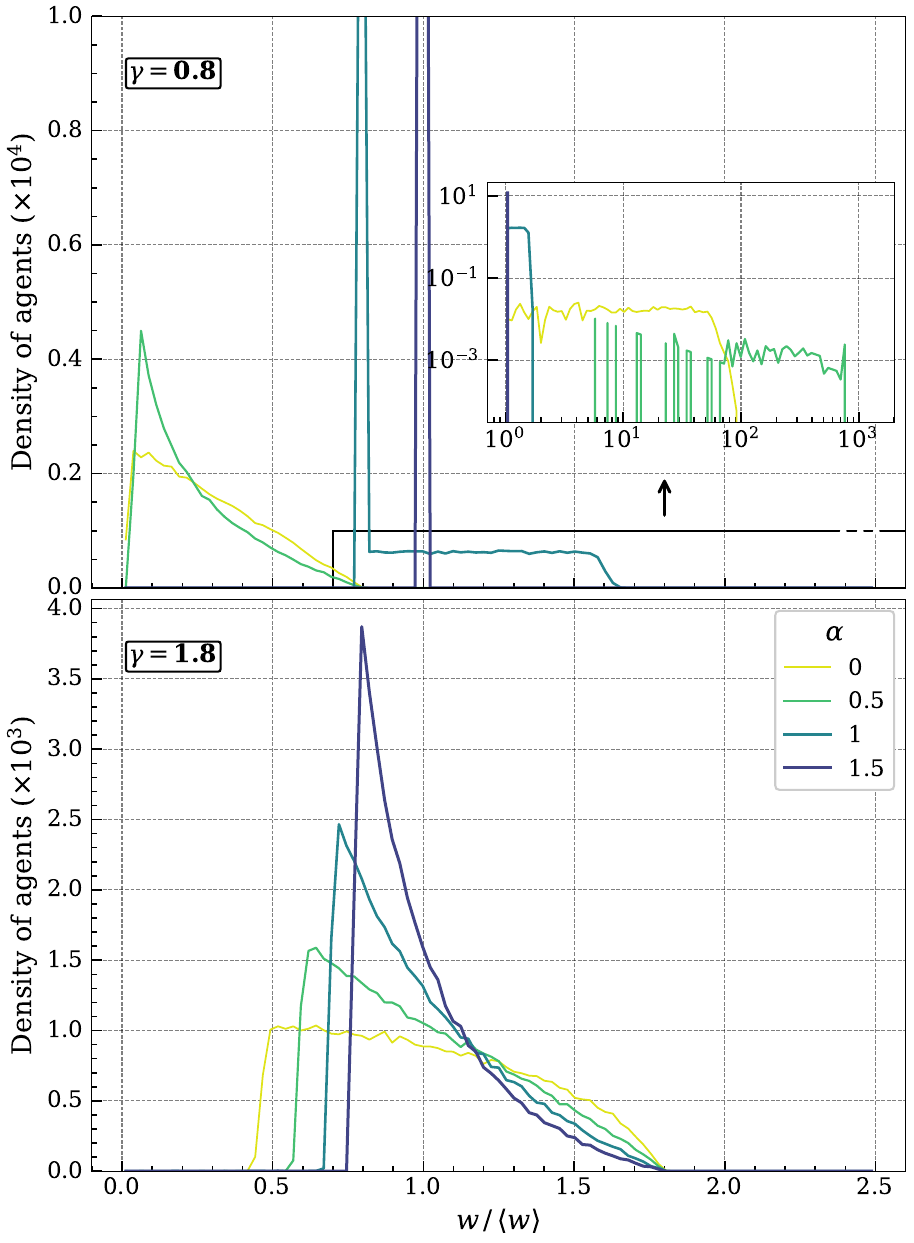}
\caption{Stationary wealth distributions for different values of $\alpha$, shown for $\gamma=0.8$ (top) and $\gamma=1.8$ (bottom). Each curve corresponds to the average over 100 independent realizations. The top panel includes an inset displaying the tail of the distribution in log--log scale with logarithmic binning, covering a wider range of wealth values than the main plot. Note that the maximum of the curves for $\alpha = 1$ and $\alpha = 1.5$ in the upper panel lies beyond the vertical scale of the plot, indicating a higher peak than shown.}

\label{fig:distrAlphas}
\end{figure}

To better understand the microscopic structure of these regimes, Fig.~\ref{fig:distrAlphas} shows the stationary wealth distributions for several values of $\alpha$ at two representative values of $\gamma$, obtained from 100 independent realizations in each case. For $\gamma=0.8$ (upper panel), regressive cases ($\alpha < 1$) display broad distributions extending down to nearly zero wealth. The inset (log-log axes) reveals the presence of extremely wealthy agents in the regressive cases. Since $\langle w \rangle = 1/N$, the upper end of the horizontal axis, $w/\langle w \rangle \sim N = 10^3$, corresponds to a single agent holding essentially the entire system wealth. In these units, single agents reach up to $10\%$ of the total wealth for $\alpha = 0$ and nearly the entire system wealth for $\alpha = 0.5$. In contrast, for $\alpha = 1$ the distribution develops a sharp lower cutoff, with a large fraction of agents accumulating near a well-defined “poverty threshold”, while the remainder of the distribution is approximately uniform up to a maximum value ($w_{\mathrm{max}} \sim 1.7\langle w \rangle$). For $\alpha = 1.5$, representative of progressive taxation, the distribution becomes sharply concentrated around the mean wealth. Only a vanishingly small fraction of agents attains lower wealth values, which are not visible on the scale of the figure. These correspond to the poorest agents, which in the stationary regime are repeatedly selected by the extremal dynamics and have their wealth continuously replaced. In practice, this typically amounts to a single agent per realization, while the rest of the population remains narrowly clustered around the mean. As a consequence, the Gini index in this regime is approximately zero, reflecting an almost perfectly egalitarian state. Although not shown here, the same qualitative behavior is observed for all $\alpha > 1$.

When $\gamma$ is increased to 1.8 (lower panel), all distributions become narrower and more homogeneous, indicating that the system approaches a more equitable regime, consistent with the decrease of the Gini index shown in Fig.~\ref{fig:giniFinalAlphas}. In addition, larger values of $\alpha$ lead to increasingly concentrated wealth distributions around the mean.

\begin{figure}[ht]
\centering
\includegraphics[width=\figwidth]{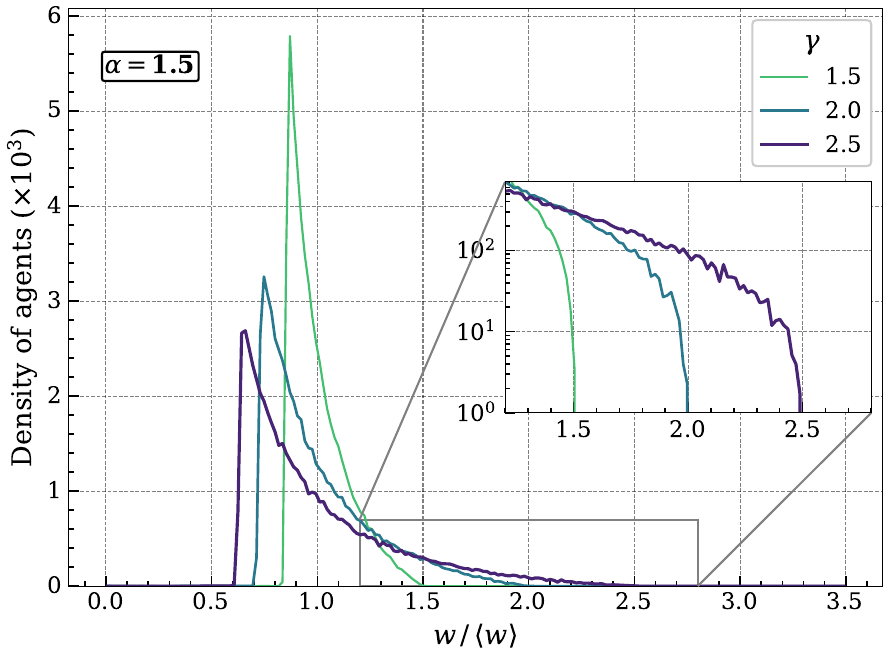}
\caption{Stationary wealth distributions for $\alpha=1.5$ and a selection of $\gamma$. The inset shows the tail of the distributions in log-linear scale.}
\label{fig:distrAlphaMayora1}
\end{figure}

Figure~\ref{fig:distrAlphaMayora1} illustrates the stationary wealth distributions for a progressive tax scheme with $\alpha = 1.5$, shown for several values of $\gamma$. All distributions exhibit a well-defined poverty threshold, corresponding to a minimum wealth value below which no agents are found. The emergence of such a threshold is not unique to the present model, but has also been reported in previous studies of the conservative model that inspired this work \cite{PIANEGONDA2003, IGLESIAS2003}. As $\gamma$ increases, the poverty threshold shifts towards lower wealth values, while the distributions broaden. This widening of the wealth distribution reflects the increase in the Gini index with $\gamma$ observed in Figs.~\ref{fig:giniFinalAlphas} and~\ref{fig:giniheatmap}. The inset displays the tails of the distributions on a log-linear scale, revealing a rapid decay at large wealth values.

An additional quantity that provides dynamical insight is the fraction of agents that have been the poorest at least once during the simulation. This indicator quantifies how many agents actively participate in the replacement dynamics and allows us to characterize the system in terms of ergodicity. In this context, we define ergodicity operationally: the dynamics is ergodic if every agent eventually has a nonzero probability of being selected as the poorest, so that over long times all agents participate in the replacement process and time and ensemble averages become equivalent. Conversely, the system is non-ergodic if a subset of agents systematically avoids replacement and remains dynamically disconnected from the extremal process.

\begin{figure}[ht]
\centering
\includegraphics[width=\figwidth]{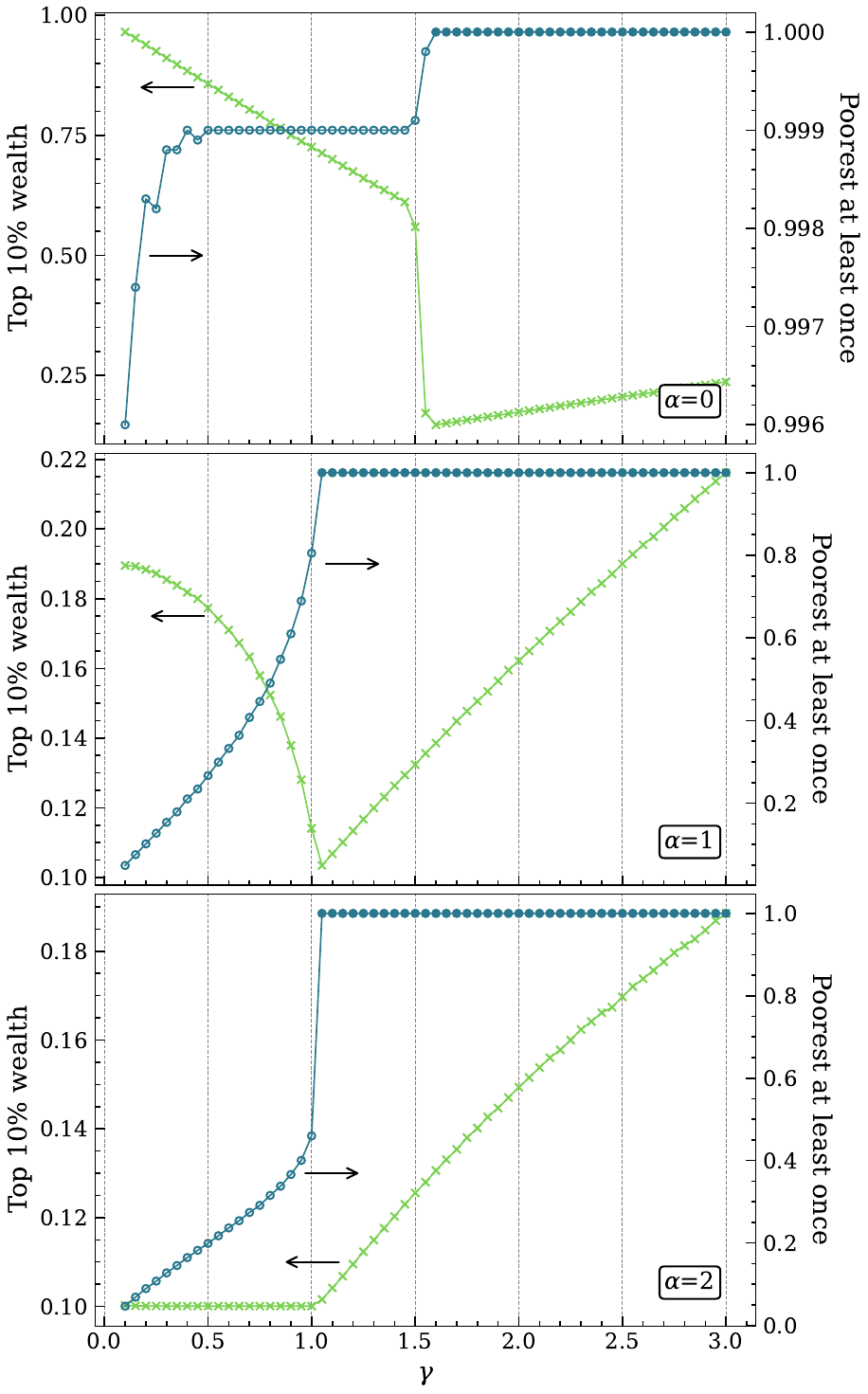}
\caption{Top 10\% wealth share and fraction of agents selected as the poorest agent at least once as a function of $\gamma$, for $\alpha=0$ (top), $\alpha=1$ (middle), and $\alpha=2$ (bottom).}
\label{fig:giniNeverPoor}
\end{figure}

A complementary measure is the Top 10\% wealth share, which captures the concentration of wealth among the richest agents. Figure~\ref{fig:giniNeverPoor} presents both quantities in three panels, corresponding to $\alpha = 0, 1,$ and $2$ (top to bottom). In each panel, the fraction of agents that have been the poorest at least once and the Top 10\% wealth share are shown as functions of $\gamma$.

Across all values of $\alpha$, a clear change in behavior occurs when the fraction of agents that have been the poorest at least once reaches unity. For values of $\gamma$ such that this fraction is smaller than one, there exist \emph{never-poor} agents who are never selected as the poorest and therefore never undergo replacement. Once the fraction becomes equal to one, the extremal process propagates through the entire population, and the system becomes fully mixing and ergodic. This change in dynamical accessibility systematically correlates with modifications in wealth concentration, as reflected in the behavior of the Top $10\%$ wealth share.

For $\alpha = 0$, the fraction of agents that have ever been the poorest remains below one over a wide range of $\gamma$, indicating the presence of a permanently protected wealthy class {\G{consisting of one or a few agents}}. In this non-ergodic regime, the Top 10\% wealth share is large, signaling strong wealth concentration. When the fraction abruptly reaches unity, the Top 10\% wealth share simultaneously drops to much lower values. Although the Gini index is not shown here, this point coincides with the discontinuous transition observed in Fig.~\ref{fig:giniFinalAlphas}, demonstrating that the macroscopic inequality jump is driven by the disappearance of dynamically shielded agents.

For $\alpha = 1$, the fraction of agents that have been the poorest at least once reaches unity close to $\gamma \simeq 1$, signaling the disappearance of never-poor agents. This point coincides with the minimum of both the Top 10\% wealth share and the Gini index, indicating that the reduction of inequality is directly associated with the onset of fully mixing dynamics.

For $\alpha=2$, the situation is qualitatively different. Although never-poor agents may still exist for $\gamma \leq 1$, progressive taxation imposes a substantially larger tax burden on wealthy agents, continuously reducing their relative wealth. As a result, even in the non-ergodic regime the wealth distribution remains tightly concentrated around the mean wealth value, yielding very low Gini values and a Top $10\%$ wealth share close to its minimal value. In other words, non-ergodicity in this case does not imply strong inequality, because the tax mechanism prevents the formation of a persistent, macroscopically dominant elite.

This behavior can be understood by examining the replacement mechanism around $\gamma \simeq 1$. Since the poorest agent receives a new wealth drawn from the interval $[0,2T)=[0,\gamma\langle w\rangle)$, the parameter $\gamma$ determines whether the replacement can lift that agent above the bulk of the distribution.
For $\gamma \le 1$, the maximum attainable wealth after replacement does not exceed the mean wealth $\langle w\rangle$. In the progressive regime, the distribution is tightly clustered around $\langle w\rangle$, forming a sharp bulk (see Fig.~\ref{fig:distrAlphas} for $\gamma=0.8$). Only a very small fraction of agents, sometimes even a single one, have wealth $w<\langle w\rangle$ and are therefore repeatedly selected as the poorest. Because the replacement cannot promote them into the bulk, the identity of the poorest agent remains essentially confined to this small subset. 
For $\gamma>1$, the replacement interval extends beyond $\langle w\rangle$, allowing the poorest agent to re-enter or even overshoot the bulk of the distribution. The identity of the poorest agent then changes over time, and the replacement mechanism propagates across the entire population. In this regime the system becomes ergodic, the wealth distribution acquires a finite and increasing width (as shown in Fig.~\ref{fig:distrAlphaMayora1}), and inequality grows as a direct consequence of the broadened distribution. This explains the increase of the Gini index for $\gamma \ge 1$. \G{In the next section, we provide a mean-field description that sheds light on this behavior from an analytical perspective.}

\subsection{\B{Mean-field theory}}
\label{subsec:meanfield}

The phenomenology described in Sec.~\ref{subsec:characterization} can be rationalized within a mean-field approximation of the dynamics. We work with the normalized dynamics introduced in Eq.~\eqref{eq:normStep}, for which the total wealth satisfies $W=\sum_i w_i=1$ after each update. It is then convenient to introduce the rescaled wealth $u_i \equiv w_i/\langle w\rangle = Nw_i$, which satisfies $\langle u\rangle = 1$ by construction. In these units the replacement value drawn at each step [Eq.~\eqref{eq:poorestReplacement}] is uniformly distributed in $[0,\gamma)$. We also introduce the rescaled time $\tau=t/N$, so that one unit of $\tau$ corresponds to $N$ microscopic steps, which is the average interval between successive replacements of a given agent.

For an agent $i$ that is not the poorest, its wealth after one update is obtained by subtracting the tax payment $T_i=w_i^\alpha T/\sum_j w_j^\alpha$ and then applying the global renormalization factor $1/(1+\Delta W)$, where $\Delta W$ follows from the change in total wealth after replacement of the poorest agent [Eq.~\eqref{eq:ExpectedTotWChange}]. In rescaled variables the exact change is
\begin{equation}
\Delta u_i = u_i(t+1)-u_i(t) = -\frac{N T_i+u_i\,\Delta W}{1+\Delta W}.
\label{eq:delta_u_exact}
\end{equation}
Both $N T_i$ and $u_i\Delta W$ are $\mathcal{O}(1/N)$ [Eq.~\eqref{eq:W(t+1)}], so the denominator can be set to $1$ up to $\mathcal{O}(N^{-2})$ corrections, and Eq.~\eqref{eq:delta_u_exact} reduces to
\begin{equation}
\Delta u_i = -\bigl(N T_i+u_i\,\Delta W\bigr)+\mathcal{O}(N^{-2}),
\label{eq:delta_u_step}
\end{equation}
where both retained terms are $\mathcal{O}(N^{-1})$ per microscopic step and accumulate to $\mathcal{O}(1)$ changes over the macroscopic time $\tau$.

The mean-field approximation replaces the two stochastic quantities by their ensemble averages: the rescaled injection $\mathrm{U}[0,\gamma)$ by its mean $\gamma/2$, and the rescaled wealth of the poorest agent $u_p=Nw_p$ by $\mathbb{E}[u_p]\equiv\mu$. Using $T=\gamma/(2N)$ and $M_\alpha\equiv\langle u^\alpha\rangle=N^{\alpha-1}\sum_jw_j^\alpha$, the rescaled tax reads $N T_i=\gamma u_i^\alpha/(2N M_\alpha)$. Substituting into Eq.~\eqref{eq:delta_u_step} and passing to continuous time ($d\tau=1/N$) yields the mean-field drift equation
\begin{equation}
\frac{du_i}{d\tau}= \mu\, u_i - \frac{\gamma\, u_i^\alpha}{2\,M_\alpha}\equiv f(u_i).
\label{eq:drift_mf}
\end{equation}

The first term is the boost received by all surviving agents from the renormalization that follows the removal of the poorest: its wealth leaves the system, lowering the total, so rescaling back to fixed $W=1$ raises every survivor's relative wealth by an amount proportional to $\mu$. The second is the deterministic drain due to taxation. Both $\mu$ and $M_\alpha$ are self-consistent functionals of the stationary distribution $\rho(u)$ and must be determined together with it.

The drift $f(u)$ has a nontrivial fixed point $f(u_*)=0$ at
\begin{equation}
    u_*^{\,\alpha-1}=\frac{2\,\mu\,M_\alpha}{\gamma},
    \label{eq:fixed_point}
\end{equation}
whose stability is governed by $f'(u_*)=\mu(1-\alpha)$: it is stable for $\alpha>1$, marginal for $\alpha=1$, and unstable for $\alpha<1$. This bifurcation at $\alpha=1$ provides a dynamical interpretation of the three qualitative regimes identified in Sec.~\ref{subsec:characterization}. For $\alpha>1$ all bulk agents are attracted toward $u_*$ and the stationary distribution is unimodal. For $\alpha<1$, on the contrary, agents above $u_*$ experience a positive drift and are never driven back to the bulk: they accumulate wealth indefinitely, which is the microscopic origin of the condensed high-inequality phase found for small $\gamma$ in Sec.~\ref{subsec:characterization}. We first solve for the stationary distribution of the ergodic bulk and validate it against simulations; its consequences for the phase transition are developed in the following sections.

The mean-field density derived below describes the ergodic bulk, which is made of those agents which cycle through the extremal dynamics, repeatedly refreshed by replacements and eventually removed as the poorest. In the ergodic phase this is the full stationary distribution; in the condensed phase the never-poor form a separate population outside the bulk. The bulk density $\rho(u)$ is supported on a compact interval $[u_a,\gamma]$: the upper edge $\gamma$ is fixed by the injection range, and the lower edge $u_a$ follows from self-consistency and will be discussed later. In steady state the probability current $J(u)=f(u)\,\rho(u)$ obeys the continuity equation
\begin{equation}
    \frac{dJ}{du}=\frac{1}{\gamma} - \frac{\gamma-u_a}{\gamma}\,\delta(u-u_a),
    \qquad u\in(u_a,\gamma),
    \label{eq:continuity}
\end{equation}
where $1/\gamma$ is the effective injection-rate density on the support (injections that fall below $u_a$ become the instantaneous new minimum and are removed at the following step, leaving the density on $(u_a,\gamma)$ unaffected) and the delta function accounts for the removal of the minimum agent, which sits deterministically at $u_a$ in the mean-field limit. Integrating Eq.~\eqref{eq:continuity} from $u$ to $\gamma$ with the boundary condition $J(\gamma)=0$ (no current crosses the injection ceiling) gives
\begin{equation}
    f(u)\,\rho(u)=\frac{u-\gamma}{\gamma},
    \qquad u\in(u_a,\gamma).
    \label{eq:flux_balance}
\end{equation}
The right-hand side is strictly negative for all $u<\gamma$, so, with $\rho\geq0$, the drift obeys $f(u)<0$ throughout the support, and
\begin{equation}
    \rho(u)=\frac{\gamma-u}{\gamma\,|f(u)|}.
    \label{eq:rho_general}
\end{equation}

Equation~\eqref{eq:rho_general} contains three unknowns, $\mu$, $u_a$ and $M_\alpha$, fixed by the self-consistency conditions
\begin{align}
    \int_{u_a}^{\gamma}\rho(u)\,du           &= 1, \label{eq:norm}   \\[3pt]
    \int_{u_a}^{\gamma}u\,\rho(u)\,du        &= 1, \label{eq:mean}   \\[3pt]
    \int_{u_a}^{\gamma}u^\alpha\,\rho(u)\,du &= M_\alpha, \label{eq:Malpha}
\end{align}
which impose normalization, the constraint $\langle u\rangle=1$, and the self-consistent definition of $M_\alpha$, respectively. For general $\alpha$ this system is solved numerically; for $\alpha=1$ and $\alpha=0$ the integrals close in elementary form. In each case the requirement $f<0$ on the support fixes its position relative to the fixed point $u_*$, as detailed below.

For $\alpha=1$, $M_1=\langle u\rangle=1$ identically, so Eq.~\eqref{eq:Malpha} is automatically satisfied and two unknowns remain. Here $f(u)=u(\mu-\gamma/2)$ has a single sign on the support, so $f<0$ simply requires $\mu<\gamma/2$, and $\rho(u)\propto(\gamma-u)/u$ on $[u_a,\gamma]$. Combining Eqs.~\eqref{eq:norm}--\eqref{eq:mean} yields the transcendental equation
\begin{equation}
    \gamma=\frac{2\bigl[\ln(\gamma/u_a)-(1-u_a/\gamma)\bigr]}
                {(1-u_a/\gamma)^2},
    \label{eq:trans_alpha1}
\end{equation}
which fixes $u_a$ as a function of $\gamma$ alone ($\mu$ follows from normalization). A real solution exists only for $\gamma\geq1$. This is not an artifact of the closed form but a universal requirement: a distribution with $\langle u\rangle=1$ supported in $[u_a,\gamma]$ can exist only if the injection ceiling exceeds the mean, $\gamma\geq1$.

In the case $\alpha=0$, $M_0=1$ exactly, and with $|f(u)|=\gamma/2-\mu u$ the system again closes in two unknowns. The support now lies below the unstable fixed point $u_*=\gamma/(2\mu)$. Setting $x\equiv u_a/\gamma$, the mean condition gives the algebraic relation
\begin{equation}
    2\mu=\gamma\,x(2-x),
    \label{eq:mu_alpha0}
\end{equation}
while normalization yields the transcendental equation
\begin{equation}
    (1-x)+\frac{1-\gamma x(2-x)}{\gamma x(2-x)}
      \ln\frac{1-\gamma x(2-x)}{1-\gamma x^2(2-x)}
    =\frac{\gamma x(2-x)}{2}.
    \label{eq:trans_alpha0}
\end{equation}
Equations~\eqref{eq:mu_alpha0}--\eqref{eq:trans_alpha0} determine $(\mu,u_a)$ as a function of $\gamma$, closing the $\alpha=0$ case.

For $\alpha > 1$ the fixed point is stable, so $f > 0$ below $u_*$ and $f < 0$ above it. The flux balance of Eq.~\eqref{eq:flux_balance} requires $f < 0$ on the support, so the bulk lies above the fixed point, $u_a > u_*$. In contrast to the $\alpha<1$ case, here $u_*$ is a stable attractor below the support: agents are driven toward $u_*$ and the stationary distribution is correspondingly concentrated. The self-consistency conditions \eqref{eq:norm}--\eqref{eq:Malpha} are then solved numerically. As above, the ergodic bulk requires $\gamma>1$; below this threshold the stable fixed point pulls all survivors toward the mean and the only stationary state is the degenerate $\rho(u)=\delta(u-1)$, consistent with the near-zero Gini values and the near-delta distributions for $\alpha>1$ at $\gamma\lesssim1$ in Sec.~\ref{subsec:characterization} (Figs.~\ref{fig:giniFinalAlphas} and~\ref{fig:distrAlphas}). For $\gamma>1$ the support opens to $[u_a,\gamma]$, reproducing the sharply peaked distributions of Fig.~\ref{fig:distrAlphaMayora1} (shown there for $\alpha=2$).

\begin{figure}[ht]
\centering
\includegraphics[width=\figwidth]{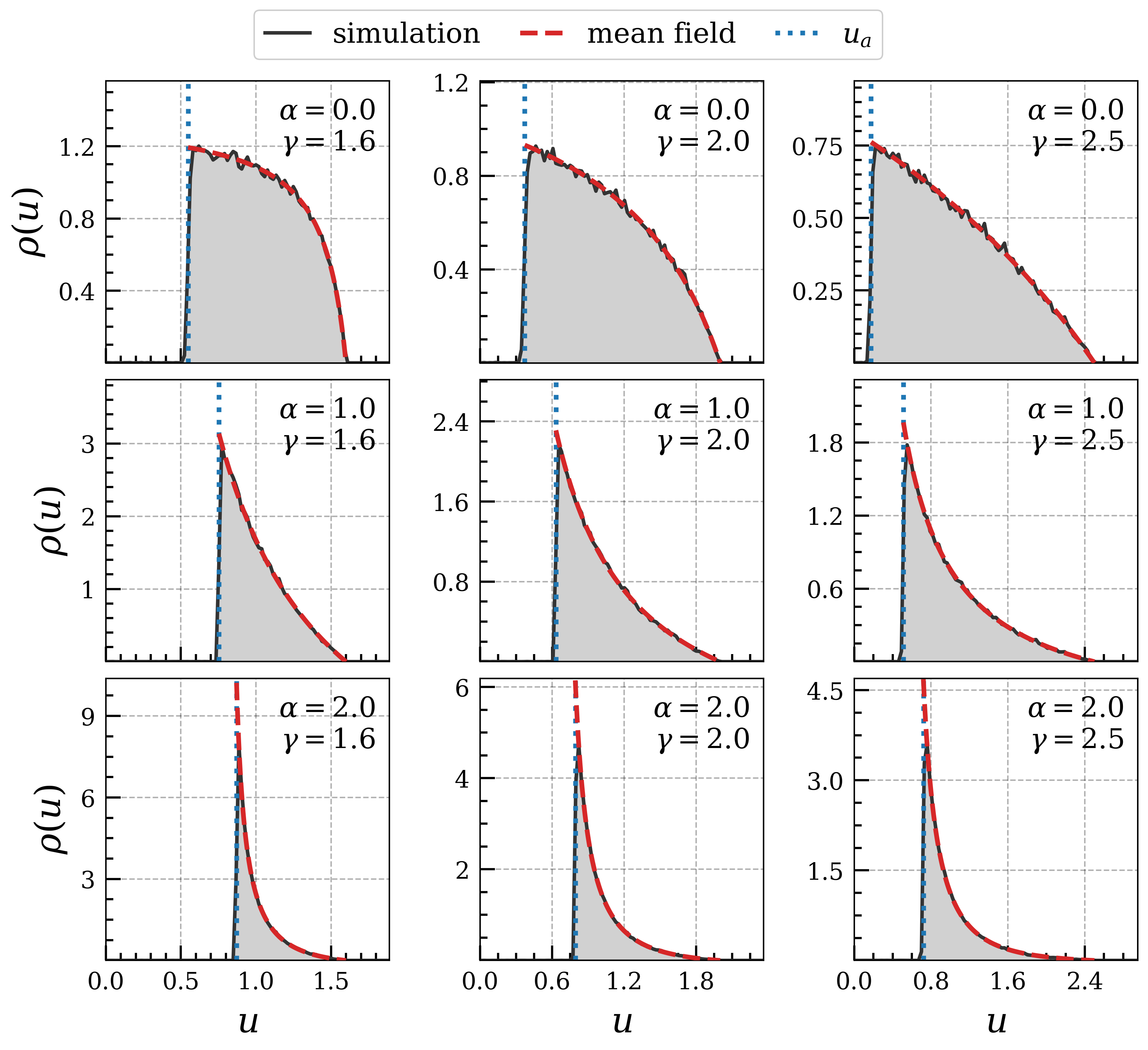}
\caption{%
    Stationary rescaled-wealth distributions $\rho(u)$ for representative values of $\alpha$ and $\gamma$. Black lines: numerical simulations (with shaded grey area underneath); red dashed lines: mean-field prediction from Eq.~\eqref{eq:rho_general} with $\mu$, $u_a$ and $M_\alpha$ obtained self-consistently from Eqs.~\eqref{eq:norm}--\eqref{eq:Malpha}. Dotted vertical lines mark the lower support edge $u_a$.
}
\label{fig:distrFit}
\end{figure}

The stationary distributions predicted by Eq.~\eqref{eq:rho_general}, with $\mu$, $u_a$ and $M_\alpha$ obtained self-consistently from Eqs.~\eqref{eq:norm}--\eqref{eq:Malpha}, are compared with simulation data in Fig.~\ref{fig:distrFit}. The agreement is quantitative and entirely parameter-free: without any fitting, the theory reproduces the position of the sharp lower edge at $u_a$ and the bulk distribution across all three regimes and a wide range of $\gamma$.

The mean-field solution clarifies several structural features of the model that are not transparent from the microscopic rules. The lower edge $u_a$ of the support is not imposed externally but emerges from the self-consistency conditions \eqref{eq:norm}--\eqref{eq:Malpha}, and matches the lower edge of the empirical bulk. For $\alpha\geq1$, moreover, the support cannot extend to the origin: there the drift vanishes linearly, $|f(u)|\propto u$, so that $\rho(u)\propto 1/u$ is non-integrable, and normalization enforces a strictly positive lower edge $u_a>0$. In all cases, the quantity $u_a\langle w\rangle$ is the minimum wealth sustained in the stationary bulk, which resembles a poverty floor analogous to that found in earlier wealth-redistribution models~\cite{PIANEGONDA2003,IGLESIAS2003}, here emerging self-consistently from the balance between taxation and the renormalization boost.

The upper edge $\gamma$ of the support is a direct consequence of the injection mechanism: since replaced agents are drawn uniformly from $[0,\gamma)$, no agent can enter with $u>\gamma$, and the drift keeps the bulk below this ceiling in the ergodic phase. Together, $u_a$ and $\gamma$ delimit the full extent of the bulk distribution with no free parameters.

Finally, for $\alpha<1$ the unstable character of $u_*$ gives a microscopic picture of how wealth condensation sets in. An agent above $u_*$ feels a positive drift, $f(u)>0$, and moves away from the bulk without ever returning; such agents make up the never-poor class, which holds a finite share of the total wealth while forming a vanishing fraction of the population. Starting from a non-condensed configuration, in which no agent lies above $u_*$, a condensate can be seeded only if a replacement is placed above $u_*$; since replacements are drawn from $[0,\gamma)$, this is possible only when the injection ceiling exceeds the fixed point, $\gamma>u_*$. Conversely, for $\gamma\leq u_*$ no replacement can overshoot $u_*$, no agent escapes, and a non-condensed configuration remains ergodic. the onset of condensation from the ergodic phase is therefore the value of $\gamma$ at which the self-consistent ergodic solution satisfies $u_*=\gamma$.


\subsection{Evidence of a phase transition}
\label{subsec:phaseTransition}

\begin{figure}[ht]
\centering
\includegraphics[width=\figwidth]{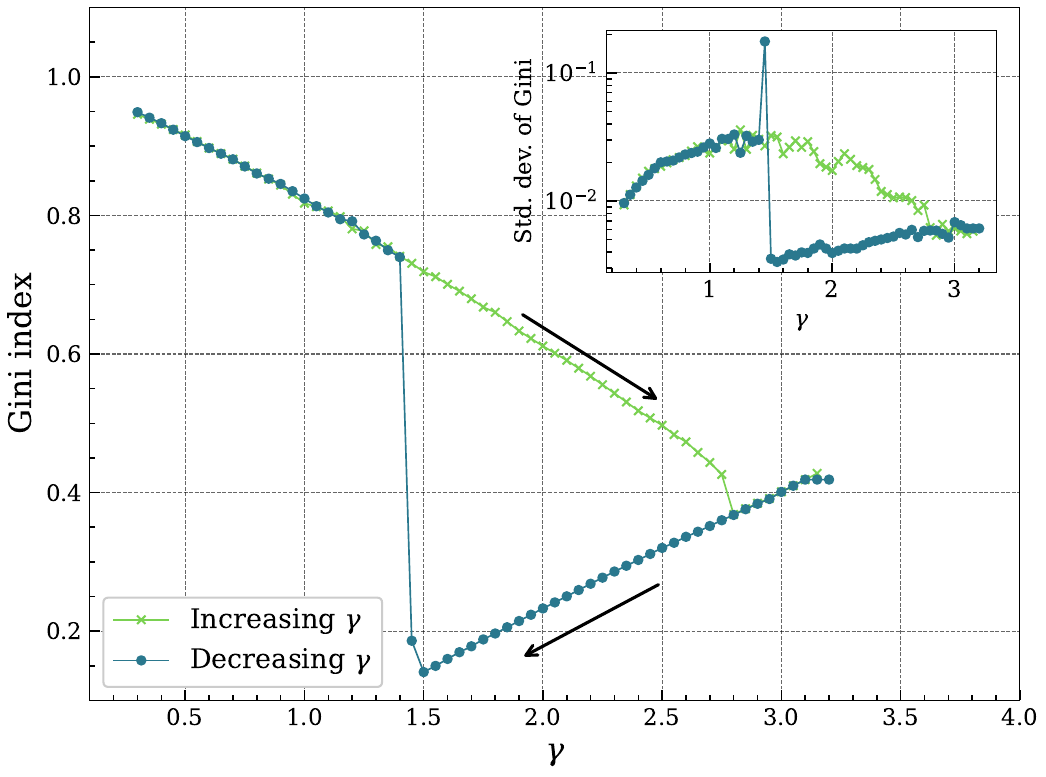}
\caption{Mean value of the stationary Gini index for the hysteresis process with varying $\gamma$ and fixed $\alpha=0$, averaged over 100 realizations. The inset shows the standard deviation for the process.}
\label{fig:histGini}
\end{figure}

The abrupt change in the Gini index for regressive taxes shown in Figs.~\ref{fig:giniFinalAlphas} and \ref{fig:giniheatmap} suggests the presence of a phase transition between two distinct stationary regimes.
To probe this hypothesis, we performed a hysteresis experiment by adiabatically increasing and then decreasing $\gamma$ for fixed $\alpha=0$, using the stationary distribution obtained at each step as the initial condition for the next run. Fig.~\ref{fig:histGini} shows the Gini index for this process, where each point represents an average over 100 realizations.

The resulting loop exhibits clear hysteresis: the transition from the high- to the low-inequality phase and back does not occur at the same $\gamma$ value. This behavior, together with the large fluctuations of the Gini index observed near the transition (shown in the inset of Fig.~\ref{fig:histGini}), is characteristic of a first-order-\G{like} phase transition. 
In the bistable region, the long-time stationary level of inequality is not uniquely determined by $(\alpha,\gamma)$: even when starting from statistically identical initial conditions (up to stochastic realizations), the system can evolve toward two distinct stationary Gini values at the same parameter point, consistent with bistability (see Appendix~\ref{app:bistability} for more details).

\begin{figure}[ht]
\centering
\includegraphics[width=\figwidth]{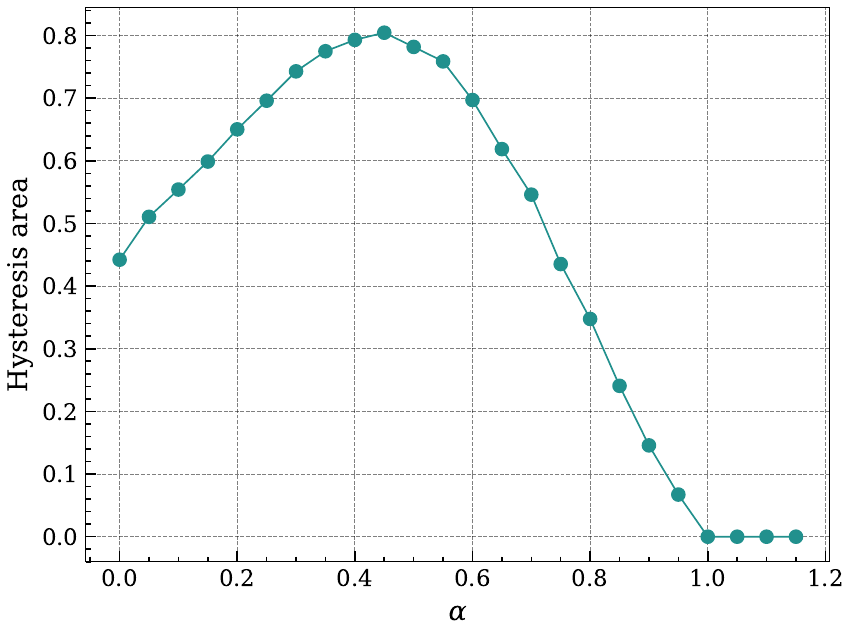}
\caption{Hysteresis area as a function of $\alpha$ in arbitrary units of $[\textrm{Gini}]\times[\gamma]$, which is effectively dimensionless.}
\label{fig:areaHistgamma}
\end{figure}

The width of the hysteresis loop quantifies the range of $\gamma$ over which both states remain dynamically stable and can be interpreted as a measure of the persistence or resilience of the unequal regime: wider loops indicate that the system remains trapped in the high-inequality phase over a larger range of conditions, reflecting stronger resistance to redistribution.
Interestingly, similar hysteresis loops are observed for other values of $\alpha<1$, indicating that this behavior is robust in all regressive tax regimes. The hysteresis area as a function of $\alpha$ is shown in Fig.~\ref{fig:areaHistgamma}. 
Remarkably, this area is not a monotonic function of $\alpha$. For strongly regressive taxation (small $\alpha$), the system is effectively frozen in a highly unequal configuration dominated by wealth condensation. In this regime, the unequal phase is extremely stable, and varying $\gamma$ produces only minor changes in the macroscopic state, resulting in a relatively narrow hysteresis loop.
As $\alpha$ increases, redistribution becomes more effective and competes with wealth condensation. At intermediate values of $\alpha$, two dynamically stable configurations are accessible: a non-ergodic state characterized by a stable wealthy elite and an ergodic state in which agents' wealth are continuously replaced. Transitions between these states require overcoming significant dynamical barriers, since both the dismantling of an established elite and its formation from a homogeneous state are hindered. Consequently, the irreversibility under cyclic variations of $\gamma$ is maximized, leading to the largest hysteresis area.
For values of $\alpha$ approaching unity from below, corresponding to taxation schemes that are still regressive but increasingly less so, redistribution becomes more effective and the high-inequality state loses stability. The system then responds almost reversibly to changes in $\gamma$, and the hysteresis loop shrinks again until finally, for $\alpha \geq 1$, there is no hysteresis.

This non-monotonic dependence of the hysteresis area on $\alpha$ is consistent with the interpretation of the transition as first-order–like, where metastable states separated by dynamical barriers control the macroscopic response of the system, as commonly observed in discontinuous transitions \cite{Binder1987}.

To further characterize the transition, we study the \B{Shannon} entropy of the \B{normalized wealth} distribution, defined as
\begin{equation}
    H = -\sum_i p_i \log_{10}(p_i),
\end{equation}
where $p_i = \frac{w_i}{W}$. In this context, $H$ quantifies the heterogeneity \B{of wealth allocation across agents}: high entropy corresponds to a more homogeneous wealth distribution, whereas low entropy indicates that wealth is concentrated among fewer agents.

Another useful tool is the Binder cumulant, defined as
\begin{equation}
    U_L = 1 - \frac{\langle w^4\rangle}{3 {\langle w^2\rangle}^2}.
\end{equation}
Originally introduced in statistical physics \cite{Binder1981}, the Binder cumulant is sensitive to higher-order moments of the distribution and is widely used to identify the nature of phase transitions. In particular, negative dips or discontinuities in $U_L$ are well-known signatures of first-order transitions. In econophysics, $U_L$ could capture the emergence of rare, extremely wealthy agents: when the distribution becomes strongly skewed, the fourth moment $\langle w^4 \rangle$ increases faster than $\langle w^2 \rangle^2$, producing a negative contribution to $U_L$.

\begin{figure}[ht]
\centering
\includegraphics[width=\figwidth]{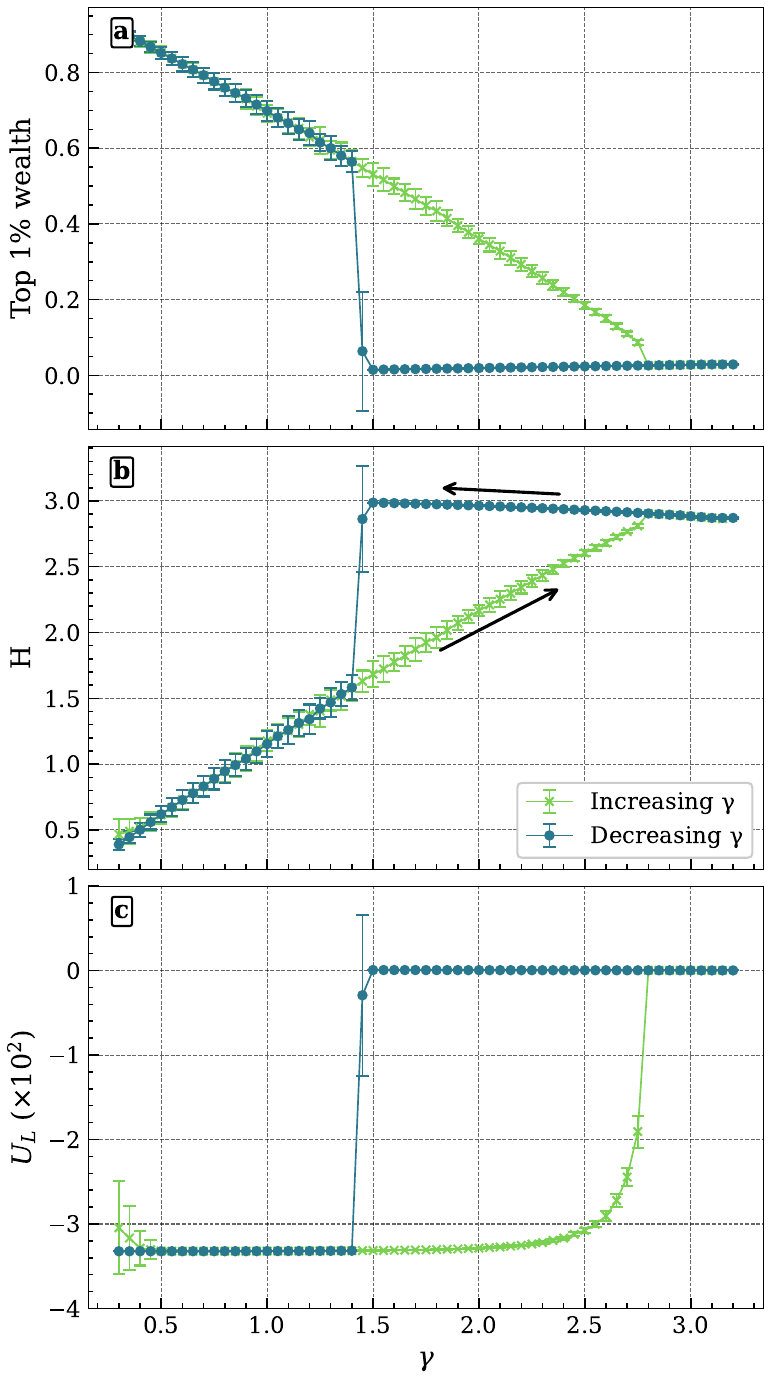}
\caption{{\textbf{a)}} Fraction of total wealth held by the Top 1\% richest agents as a function of $\gamma$ for the hysteresis process, with $\alpha=0$. \textbf{b)} Entropy of the distribution as a function of $\gamma$. \textbf{c)} Binder Cumulant as a function of $\gamma$.}
\label{fig:histEntropy}
\end{figure}

Using these two complementary observables, Fig.~\ref{fig:histEntropy} extends the hysteresis analysis beyond the Gini index by examining the Top $1\%$ wealth share, the entropy $H$, and the Binder cumulant $U_L$ for a regressive taxation scheme ($\alpha=0$). As in the case of the Gini index, the stationary values of all these observables depend on whether $\gamma$ is increased or decreased quasistatically, giving rise to distinct increasing and decreasing branches.

For sufficiently large values of $\gamma$ ($\gamma \gtrsim 2.8$), the increasing and decreasing branches collapse onto a single curve for all observables, indicating the absence of hysteresis. In this regime, redistribution dominates the dynamics and the system converges to a unique stationary state characterized by low wealth concentration, high entropy, and Binder cumulant values close to zero, consistent with weak, approximately Gaussian fluctuations.

\G{For intermediate values of $\gamma$ ($1.5 \lesssim \gamma \lesssim 2.8$), the increasing and decreasing branches separate, signaling the presence of hysteresis over the same $\gamma$ interval identified in the Gini index (Fig.~\ref{fig:histGini}). For even smaller values of $\gamma$, the branches merge again within numerical fluctuations. This behavior indicates that the bistable region is confined to an intermediate range of $\gamma$, while outside this range the system evolves toward a unique stationary state regardless of the direction in which $\gamma$ is varied.}

\G{The Top $1\%$ wealth share shown in Fig.~\ref{fig:histEntropy}\textbf{a} follows the same qualitative pattern observed for the Gini index. Within the bistable region, the increasing and decreasing branches remain clearly separated, indicating that the coexistence of stationary states is accompanied by markedly different levels of wealth concentration among the richest agents. In particular, along the increasing branch the wealth share of the richest $1\%$ is systematically larger, showing that a substantial fraction of the system's wealth becomes concentrated in a very small number of individuals. By contrast, along the decreasing branch the Top $1\%$ wealth share remains close to zero throughout most of the bistable region, indicating the absence of a dominant wealthy elite and a much more broadly distributed allocation of wealth. The coexistence of these two branches therefore reflects two qualitatively different stationary organizations of wealth under identical values of $\gamma$. This branch dependence highlights that the hysteresis is not only reflected in global inequality measures, but also in the degree of wealth accumulation by the economic elite.}

\G{The entropy (Fig.~\ref{fig:histEntropy}\textbf{b}) and the Binder cumulant (Fig.~\ref{fig:histEntropy}\textbf{c}) provide complementary information about the internal structure of the coexisting stationary states. The entropy remains systematically lower along the increasing branch, indicating that, for the same value of $\gamma$, wealth is distributed among a smaller effective fraction of the population than along the decreasing branch. The Binder cumulant exhibits an equally pronounced branch dependence. In particular, the strongly negative values observed along part of the increasing branch indicate highly skewed wealth distributions dominated by rare but exceptionally wealthy agents.}

\G{Taken together, these results indicate that the observed hysteresis is a robust property of the system. Although the different observables probe distinct aspects of the wealth distribution, they consistently identify the same bistable region. Moreover, they show that the two coexisting stationary states differ in the concentration of wealth at the top of the distribution, in the effective number of agents holding significant wealth, and in the prevalence of extreme wealth fluctuations.}


\subsection{\B{Spinodal structure of the transition}}
\label{subsec:spinodals}

The hysteresis loop and its non-monotonic area, observed numerically in Sec.~\ref{subsec:phaseTransition}, admit a quantitative explanation within the mean-field theory of Sec.~\ref{subsec:meanfield}. 

As illustrated in Fig.~\ref{fig:histGini}, the hysteresis loop defines a range of the control parameter $\gamma$ in which two stationary states are locally stable. The two branches terminate at different values of $\gamma$: along the increasing-$\gamma$ branch, the high-inequality state persists until it loses stability and collapses, whereas along the decreasing-$\gamma$ branch, the low-inequality state persists until it becomes unstable and the system jumps to the high-inequality state. These two stability limits define the \emph{upper} and \emph{lower spinodals}, respectively.

The upper spinodal $\gamma_c^{\rm up}(\alpha)$ is therefore the largest value of $\gamma$ for which the high-inequality state remains stable, while the lower spinodal $\gamma_c^{\rm lo}(\alpha)$ is the smallest value of $\gamma$ for which the low-inequality state remains stable. Between these two spinodals, both states are locally stable, giving rise to bistability and hysteresis. Both spinodals can be obtained from the mean-field drift \eqref{eq:drift_mf}. In what follows, we derive their dependence on $\alpha$ and compare the resulting predictions with the values measured along the corresponding branches of the hysteresis loop.

\begin{figure}[ht]
\centering
\includegraphics[width=\figwidth]{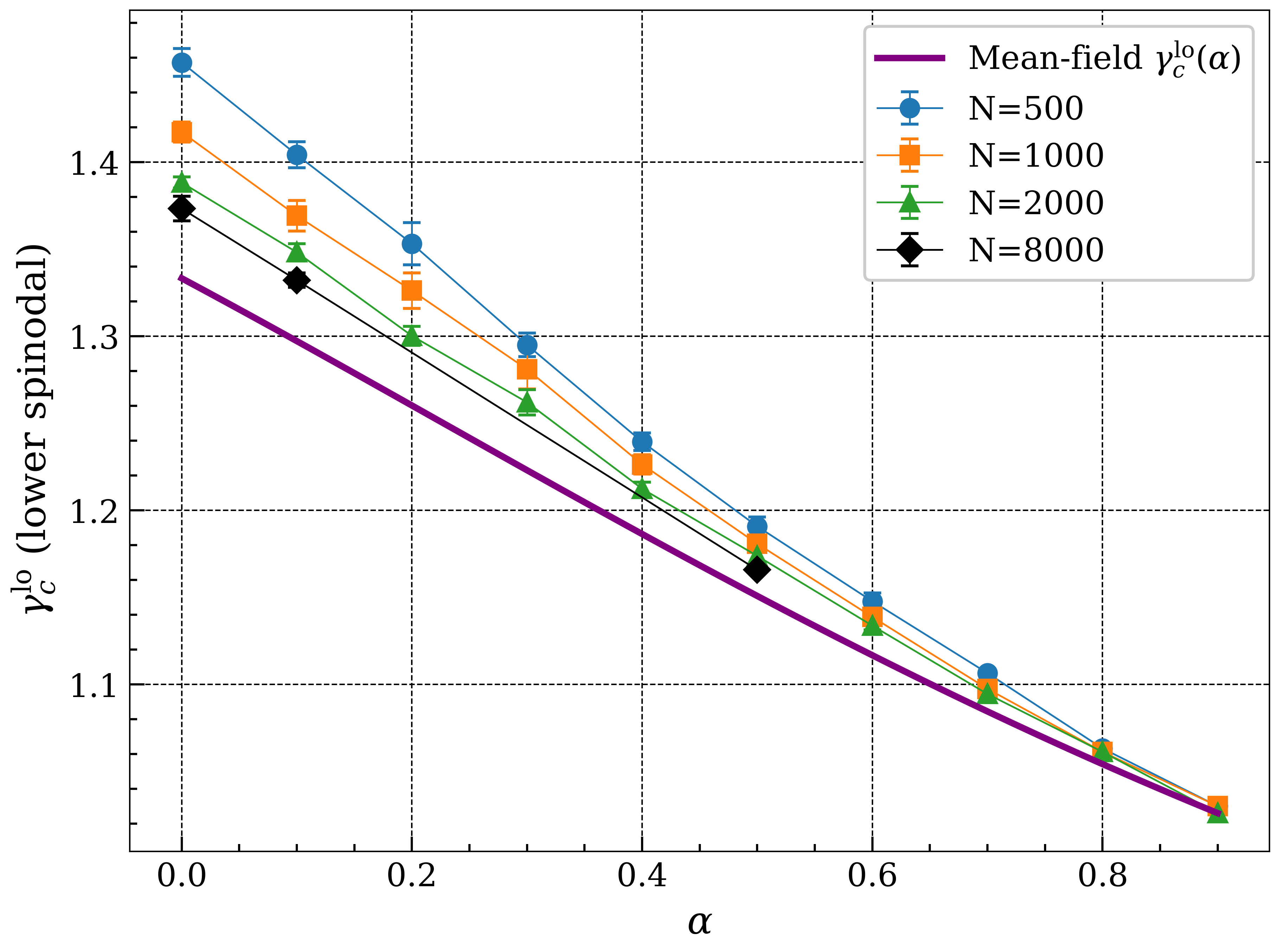}
\caption{%
    Measured lower spinodal $\gamma_c^{\rm lo}$ as a function of $\alpha$ for several system sizes $N$ (symbols), obtained from the decreasing-$\gamma$ branch of the hysteresis protocol, together with the mean-field prediction of Eq.~\eqref{eq:lower_spinodal} (solid line).
}
\label{fig:lowerspinodalN}
\end{figure}

\paragraph{Lower spinodal.}
The ergodic bulk loses stability at the condensation-onset boundary $u_*=\gamma$ anticipated in Sec.~\ref{subsec:meanfield}, where the unstable fixed point $u_*$ of the drift [Eq.~\eqref{eq:drift_mf}] reaches the upper edge of the injection range. For $\alpha<1$ the flux constraint requires $u_*\geq\gamma$; the boundary case
\begin{equation}
    u_*\bigl(\gamma_c^{\rm lo},\,\mu,\,M_\alpha\bigr)=\gamma_c^{\rm lo}
    \label{eq:lower_spinodal}
\end{equation}
marks the smallest $\gamma$ for which a self-consistent ergodic distribution exists. Below it, replacements can be injected above $u_*$, where the drift is positive, so that agents escape upward and feed the condensate. Using Eqs.~\eqref{eq:fixed_point} and \eqref{eq:norm}--\eqref{eq:Malpha}, condition \eqref{eq:lower_spinodal} fixes $\gamma_c^{\rm lo}(\alpha)$ self-consistently; for $\alpha=0$ it gives $\gamma_c^{\rm lo}=4/3$, and it decreases smoothly with $\alpha$. Because it is set by the intensive mean-field densities $\mu$ and $M_\alpha$, $\gamma_c^{\rm lo}$ is a property of the $N\to\infty$ limit. The lower spinodal measured along the decreasing-$\gamma$ branch carries a finite-size correction that decreases with $N$: as shown in Fig.~\ref{fig:lowerspinodalN}, the measured values approach $\gamma_c^{\rm lo}(\alpha)$ from above across the whole range of $\alpha$. We stress that this size dependence concerns the location of the stability boundary; it is consistent with the size-independence of the stationary inequality within each phase established in Appendix~\ref{app:finitesize}, since the latter is an intensive bulk property whereas the spinodal is a threshold that sharpens with $N$.

\paragraph{Upper spinodal.}

In the condensed phase one or a few wealthy agents hold a finite share $\beta$ of the total wealth $W$, so the remaining agents form an ergodic bulk that shares only the fraction $1-\beta$; their mean wealth is thus a factor $1-\beta$ below the global mean $\langle w\rangle$. The poorest agent, however, is still replaced by a draw up to $\gamma\langle w\rangle$, set by that global mean, so relative to the impoverished bulk the injection is a factor $1/(1-\beta)$ larger: the bulk behaves as an ergodic phase at an effective redistribution strength $\gamma_{\rm eff}=\gamma/(1-\beta)>\gamma$.

The high-inequality state survives only while the boost $\mu(\gamma_{\rm eff})$ outweighs the tax paid by the rich agents. This boost weakens as $\gamma$ grows: $\mu$ is the rescaled wealth of the poorest agent, and a stronger redistribution widens the bulk while its mean stays pinned at unity, so its lower edge (and hence $\mu$) is pushed toward zero. When the boost can no longer cover the tax, the high-inequality state gives way; this is the upper spinodal.

As $N\to\infty$ we find that the fraction of wealth held by the rich agent at the spinodal vanishes, $\beta\to0$, even though the agent itself remains macroscopically wealthy, its rescaled wealth diverging as $u_c=\beta N\to\infty$ (Appendix~\ref{app:saddlenode}). The two limits enter the balance in complementary ways: $\beta\to0$ leaves the bulk unperturbed,
$\gamma_{\rm eff}=\gamma/(1-\beta)\to\gamma$, while $\beta N\to\infty$ sends the regressive per-unit tax $(\beta N)^{\alpha-1}\to0$. With no tax left to offset, the balance reduces to the vanishing of the bulk boost,
\begin{equation}
    \mu\bigl(\gamma_{c,\infty}^{\rm up}\bigr)=0 .
    \label{eq:upper_spinodal_cond}
\end{equation}
Setting $\mu=0$ in Eq.~\eqref{eq:rho_general} gives the closed-form
ergodic density $\rho(u)\propto(\gamma-u)\,u^{-\alpha}$ on $[0,\gamma]$, which is normalizable for $\alpha<1$. Imposing the mean constraint \eqref{eq:mean},
\begin{equation}
    \langle u\rangle
    =\frac{\displaystyle\int_0^{\gamma} u(\gamma-u)\,u^{-\alpha}\,du}
          {\displaystyle\int_0^{\gamma} (\gamma-u)\,u^{-\alpha}\,du}
    =\frac{1-\alpha}{3-\alpha}\,\gamma
    \;=\;1 ,
\end{equation}
yields the upper spinodal in the mean-field limit,
\begin{equation}
    \;\gamma_{c,\infty}^{\rm up}(\alpha)=\frac{3-\alpha}{1-\alpha}\;
    \label{eq:upper_spinodal}
\end{equation}
which for $\alpha=0$ gives $\gamma_{c,\infty}^{\rm up}=3$.

At the finite size $N=1000$ used throughout this work the never-poor share $\beta$ is not negligible, so $\gamma_{\rm eff}>\gamma$ and the balance \eqref{eq:upper_spinodal_cond} is met at a smaller $\gamma$ than $\gamma_{c,\infty}^{\rm up}$. The measured upper spinodal therefore lies below Eq.~\eqref{eq:upper_spinodal}, the more so the larger $\alpha$, since the regressive correction $\propto N^{\alpha-1}$ grows with $\alpha$. The full finite-$N$ condition is a saddle-node of the condensate balance in $\beta$ (Appendix~\ref{app:saddlenode}); it contains no free parameters and reproduces the measured upper spinodals to within a few percent across the whole range of $\alpha$.

\begin{figure}[ht]
\centering
\includegraphics[width=\figwidth]{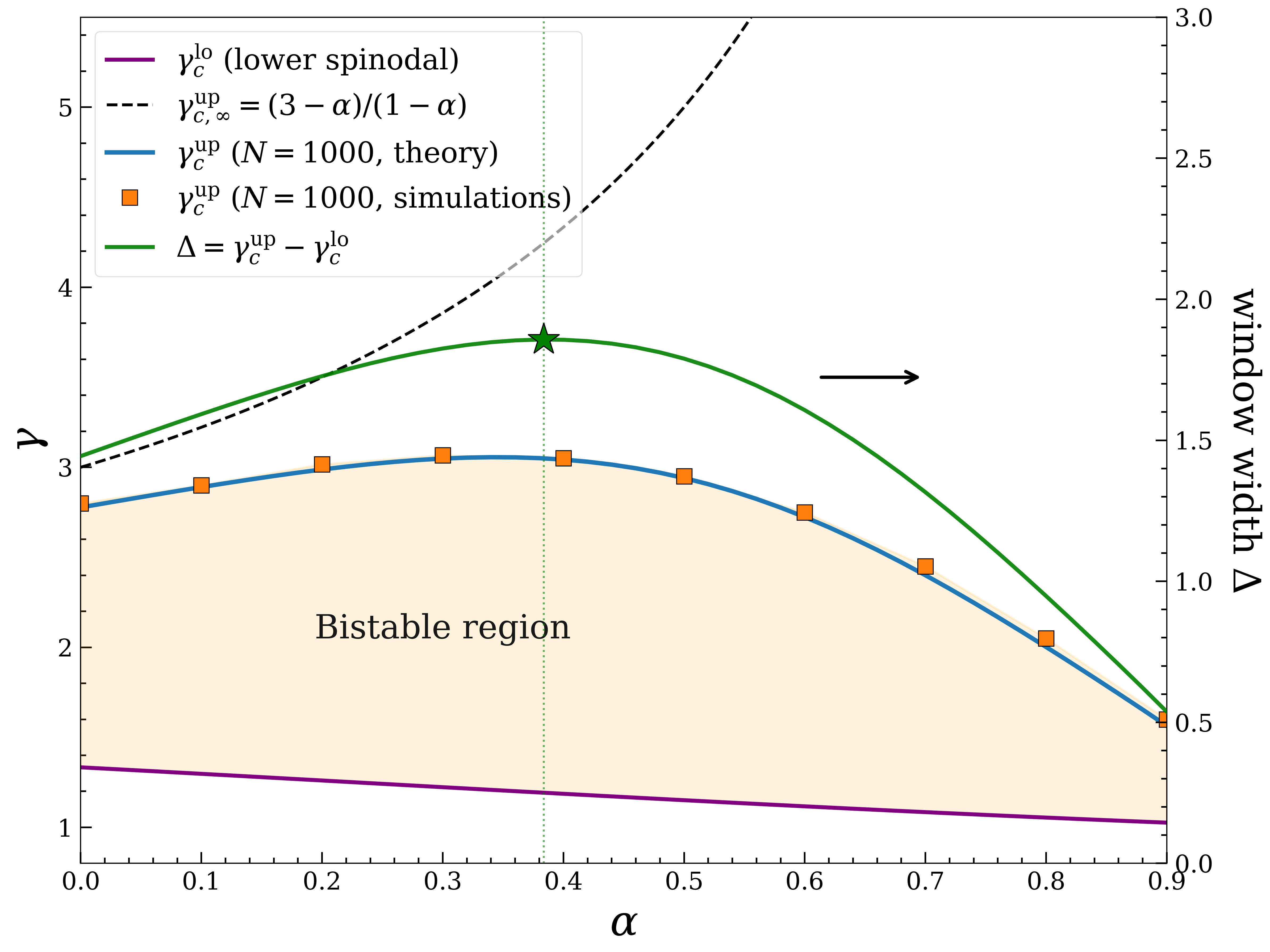}
\caption{%
    Spinodal structure of the transition as a function of the progressivity $\alpha$, for $N=1000$. Purple: lower spinodal $\gamma_c^{\rm lo}(\alpha)$ from the mean-field condition $u_*=\gamma$. Black dashed: upper spinodal in the mean-field limit, $\gamma_{c,\infty}^{\rm up}=(3-\alpha)/(1-\alpha)$. Blue: finite-$N$ upper spinodal from the saddle-node theory (Appendix~\ref{app:saddlenode}); orange squares: upper spinodal measured with the adiabatic hysteresis protocol of Fig.~\ref{fig:histGini}. The shaded band is the bistable region. Green (right axis): width of the bistable region
    $\Delta=\gamma_c^{\rm up}-\gamma_c^{\rm lo}$, whose flat maximum at $\alpha\simeq0.38$ (star) is close to the maximum of the hysteresis area in Fig.~\ref{fig:areaHistgamma}.%
}
\label{fig:spinodals}
\end{figure}

\paragraph{Bistable region and the hysteresis area.}
Fig.~\ref{fig:spinodals} collects both spinodals as functions of
$\alpha$ for $N=1000$: the lower spinodal $\gamma_c^{\rm lo}(\alpha)$ from Eq.~\eqref{eq:lower_spinodal}, the upper spinodal from the finite-$N$ theory together with its mean-field limit \eqref{eq:upper_spinodal}, and the values obtained directly from the hysteresis protocol. The agreement between theory and simulation is quantitative and parameter-free. Whereas the mean-field limit \eqref{eq:upper_spinodal} grows monotonically and diverges as $\alpha\to1^-$, the finite-$N$ upper spinodal is non-monotonic: it rises, reaches a maximum near $\alpha\approx0.34$, and then falls, because the regressive correction increasingly suppresses it as $\alpha\to1$. Since $\gamma_c^{\rm lo}(\alpha)<\gamma_{c,\infty}^{\rm up}(\alpha)$ for all $\alpha<1$, the two spinodals bound a finite bistable region in which the condensed and ergodic states coexist, while outside it only one phase is stable.

The width of the bistable region, $\Delta(\alpha)=\gamma_c^{\rm up}(\alpha)-\gamma_c^{\rm lo}(\alpha)$, measures the range of $\gamma$ over which the two states coexist and is the mean-field counterpart of the width of the hysteresis loop. As shown in Fig.~\ref{fig:spinodals}, $\Delta(\alpha)$ is also non-monotonic, with a flat maximum at $\alpha\simeq0.38$ (located from the smooth theoretical curves on a fine $\alpha$ grid). This provides a direct, quantitative explanation for the non-monotonic hysteresis area of Fig.~\ref{fig:areaHistgamma}: for strongly regressive taxation ($\alpha\to0$) the condensate is extremely robust but the coexistence interval is narrow; as $\alpha$ increases, redistribution competes with condensation and widens the bistable window; and as $\alpha\to1$ the condensed state loses stability and the window closes. The maximum of $\Delta(\alpha)$ lies close to, but need not coincide exactly with, that of the hysteresis area ($\alpha\approx0.45$ in Fig.~\ref{fig:areaHistgamma}): the width $\Delta$ measures only the $\gamma$-extent of the coexistence region, whereas the loop area also integrates the magnitude of the inequality jump across it, so the two are related but distinct measures and their maxima are expected to agree only qualitatively. The spinodal analysis thus recovers, from the mean-field drift \eqref{eq:drift_mf} alone, both the existence of two coexisting states and the non-monotonic dependence of their persistence on the taxation structure.


\section{Discussion} 
\label{sec:conclusions}

The results presented above provide a coherent picture of how taxation and redistribution jointly shape wealth dynamics in a simple non-conservative agent-based economy. By systematically exploring the parameter space, we have shown that the system self-organizes into distinct stationary regimes whose properties depend sensitively on both the tax strength and progressivity. These regimes differ not only in their overall level of inequality, but also in the microscopic organization of wealth and in the degree of economic mobility experienced by individual agents.

\B{Within this framework, the parameters act as effective controls of the taxation–replacement dynamics. The parameter $\alpha$ determines how the tax burden is distributed across agents, ranging from regressive to progressive taxation, whereas $\gamma$ controls the amount of wealth collected through taxation and subsequently injected into the replacement process. Exploring this parameter space reveals the emergence of qualitatively distinct stationary regimes, including abrupt transitions, hysteresis, bistability, and markedly different patterns of wealth concentration.}

Under regressive taxation ($\alpha<1$), the system undergoes a qualitative change as the tax strength parameter $\gamma$ is varied. For sufficiently small $\gamma$, a non-ergodic regime emerges, characterized by the formation of a class of {\G{one or a few}} {\it never-poor} agents that permanently escape replacement and retain a considerable fraction of total wealth. This regime corresponds to a ``frozen'' economic state dominated by entrenched elites, in contrast to a ``mobile'' regime in which all agents eventually renew their wealth.

As $\gamma$ increases, ergodicity is restored and all agents participate in the replacement dynamics. The wealth distribution becomes more homogeneous and inequality is reduced. The transition between these regimes is abrupt and accompanied by hysteresis, indicating the existence of two distinct stationary states over a finite range of $\gamma$. This bistable behavior is reminiscent of first-order phase transitions in physical systems, where metastable states are separated by finite barriers. In economic terms, this implies that once inequality becomes entrenched, stronger redistributive action is required to reverse it, highlighting the persistence and resilience of wealth concentration under regressive taxation. We also verified that the discontinuous drop in the stationary Gini index and the location of the transition are essentially unchanged when increasing $N$ from $500$ to $10^4$. 

The hysteretic character of the transition is consistently captured by different macroscopic observables. In addition to the Gini index, we find that the Top $1\%$ wealth share, the entropy of the wealth distribution, and the Binder cumulant all display separated increasing and decreasing branches over the same interval of $\gamma$. Although these quantities probe different aspects of the distribution, they identify a common region of bistability, underscoring the robustness of the observed transition. Importantly, entropy and the Binder cumulant provide complementary information by being sensitive to the internal structure of the wealth distribution and to the presence of extreme wealth fluctuations, rather than only to overall inequality levels.

In contrast, under progressive taxation ($\alpha \ge 1$), no abrupt transition is observed. Wealth distributions remain comparatively homogeneous, and no persistent elite capable of escaping replacement emerges. This behavior is consistent with a continuous evolution or crossover between regimes and highlights the central role of tax progressivity in sustaining economic mobility and preventing the formation of entrenched wealth hierarchies.

\G{
These numerical findings are underpinned by a mean-field theory of the dynamics. Replacing the wealth of the poorest agent and the injected value by their averages yields a deterministic drift for the rescaled wealth whose non-trivial fixed point changes stability exactly at $\alpha=1$ (attracting under progressive taxation, repelling under regressive taxation). This single bifurcation separates the qualitative regimes: a unimodal distribution concentrated around the mean for $\alpha>1$, and, for $\alpha<1$, the escape of agents beyond the unstable fixed point to form the never-poor class, with the neutral case $\alpha=1$ marking the marginal boundary between them. Solving the stationary distribution self-consistently reproduces the simulated bulk distributions across all regimes with no adjustable parameters, and clarifies the microscopic origin of features such as the self-organized lower wealth cutoff, a ``poverty line'' that emerges from the balance between taxation and the replacement dynamics.
}

\G{
The same drift accounts for the transition itself for $\alpha<1$. The ergodic phase loses stability when the unstable fixed point reaches the top of the injection range, which sets the lower edge of the coexistence region, while the high-inequality phase dissolves once redistribution can no longer sustain the wealthy agents, which sets the upper edge. These two conditions are the spinodals of the first-order-like transition, and they bound the bistable region in quantitative agreement with the values extracted from the hysteresis loops. The theory moreover predicts that the width of this region, which is a measure of how persistently inequality resists redistribution, varies non-monotonically with the progressivity parameter $\alpha$, peaking at intermediate $\alpha$, which explains the non-monotonic hysteresis area found in the simulations. A single mean-field picture thus ties together the stationary distributions, the location and order of the transition, and the resilience of wealth concentration.
}

\B{The present framework identifies generic mechanisms linking taxation structure, redistribution, and collective inequality patterns. Future extensions could explore both alternative redistribution schemes and possible connections between the effective model parameters and empirical fiscal indicators.}
\R{An interesting extension would be to replace the extremal redistribution rule by a universal basic income mechanism, in which the collected tax revenue is redistributed equally among all agents. Such a modification would allow a direct comparison between targeted redistribution toward the least-advantaged position and universal transfers, two conceptually distinct approaches that have both received attention in economic and policy discussions. Exploring how these alternative redistribution schemes affect inequality, economic mobility, and the emergence of collective regimes constitutes a promising direction for future work.}

Overall, our results demonstrate that even minimal stochastic redistribution rules can generate rich collective dynamics, including non-ergodicity, bistability, hysteresis, and abrupt changes in inequality. The emergence of {\it never-poor} agents provides a clear microscopic mechanism for persistent wealth concentration and its resistance to redistribution. 
\R{This framework offers a minimal setting in which to investigate how taxation structure and extremal replacement dynamics interact to generate qualitatively distinct inequality regimes and collective organizational states.}

\section*{Acknowledgments}
L.G. acknowledges support from ACAL through a mobility fellowship that enabled a research visit to UFRGS.
\G{SG and JRI acknowledge financial support from CNPq through projects 309560/2025-0, 304101/2020-6, and 406820/2025-2.}




\appendix
\section{Finite-size analysis}
\label{app:finitesize}

\begin{figure}[ht]
    \centering
    \includegraphics[width=\figwidth]{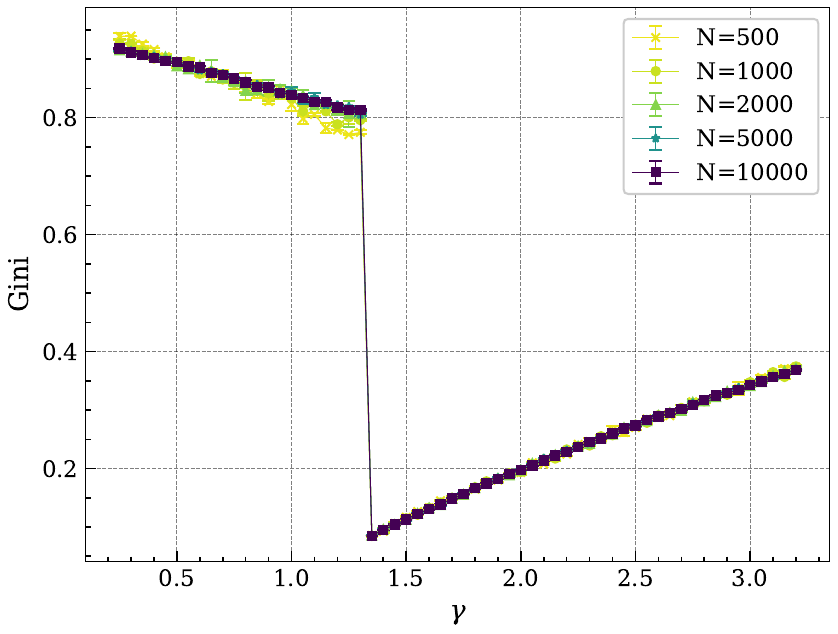}
    \caption{Stationary Gini index for $\alpha=0.5$ and varying system size.}
    \label{fig:finitesize}
\end{figure}

To assess finite-size effects, we repeated the simulations for several system sizes 
$N=\{500, 1000, 2000, 5000, 10000\}$, keeping all other parameters fixed. 
Figure~\ref{fig:finitesize} shows the stationary Gini index $G(\gamma)$ for $\alpha=0.5$.

The curves collapse across system sizes within error bars, and the location of the transition remains unchanged as $N$ increases. In particular, the position of the sharp variation in $G(\gamma)$ does not display any systematic drift with $N$, indicating that the transition is not a finite-size artifact.

\G{Moreover, the magnitude of the jump in the Gini index remains stable as $N$ increases, and both branches remain dynamically stable, suggesting that the observed bistability persists in the thermodynamic limit (the size dependence of the
spinodal boundaries that enclose the bistable region is analyzed in Appendix~\ref{app:saddlenode}). Fluctuations around the stationary states decrease with system size, as expected, but do not alter the qualitative structure of the transition.}

These results support the interpretation of the transition as a robust collective phenomenon rather than a finite-size effect.

\section{Bistability and dependence on initial conditions}
\label{app:bistability}

\subsection*{Temporal evolution and dynamical bistability}

To explicitly demonstrate the presence of dynamical bistability, we analyze the temporal evolution of the Gini index for a fixed value of the tax strength parameter $\gamma = 1.75$, and $\alpha = 0$. In this analysis, the system is not driven along the hysteresis cycle; instead, $\gamma$ is kept constant and the dynamics is allowed to relax from an ensemble of statistically equivalent initial conditions. 

\begin{figure}[ht]
\centering
\includegraphics[width=\figwidth]{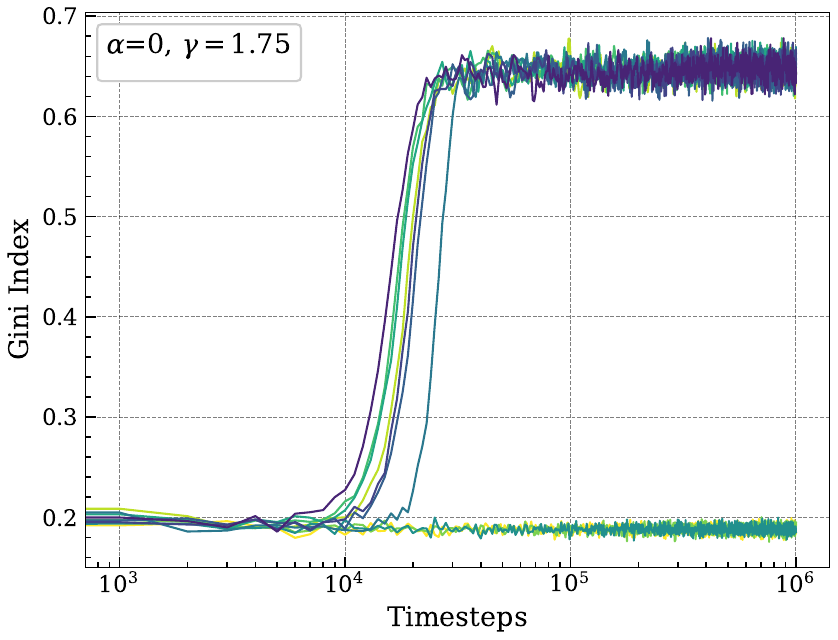}
\caption{Temporal evolution of the Gini index for $\gamma = 1.75$ and $\alpha = 0$ for $10$ independent realizations. All trajectories start from Gaussian initial wealth distributions with identical mean wealth $\langle w\rangle = \frac{1}{N}$ and standard deviation $\sigma = 0.4\langle w\rangle$, and differ only due to microscopic fluctuations.}
\label{fig:temporal_bistability}
\end{figure}

In Fig.~\ref{fig:temporal_bistability} we show the time evolution of the Gini index for $10$ independent realizations. 

In all cases, the initial wealth distribution is Gaussian, with the same mean wealth $\langle w\rangle = \frac{1}{N}$ and standard deviation $\sigma_w = 0.4\langle w\rangle$, while individual realizations differ only through microscopic fluctuations due to random sampling.
\B{The specific shape of the initial distribution is not intended to represent a realistic socioeconomic state, but rather to provide statistically controlled initial conditions from which the long-time dynamics can be analyzed.}

Despite their identical macroscopic initial conditions, the trajectories separate after a transient and converge towards two distinct stationary values of the Gini index. A subset of realizations relaxes to a low-inequality state, while the remaining ones become trapped in a high-inequality configuration.

This behavior demonstrates that, for this value of $\gamma$, the system exhibits dynamical bistability: two stationary states are accessible depending on initial microscopic fluctuations. While initial conditions sufficiently close to one of the equilibria systematically converge to that state (not shown here), intermediate configurations can evolve toward either stationary value, with microscopic fluctuations effectively selecting the long-term outcome.
\B{More generally, the results indicate that the existence of the stationary regimes themselves is robust and does not depend on the detailed form of the initial wealth distribution, except within the bistable region where fluctuations determine which branch is ultimately selected.}

\subsection*{Microscopic structure of the bistable stationary states}

\begin{figure}[ht]
\centering
\includegraphics[width=\figwidth]{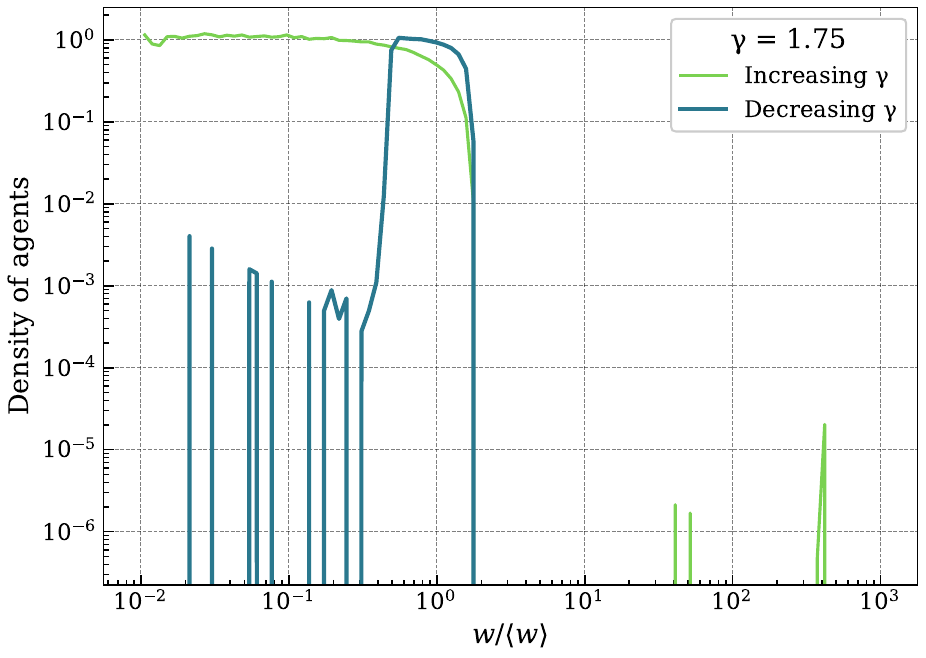}
\caption{Stationary wealth distributions in log-log scale with logarithmic binning for $\gamma = 1.75$ and $\alpha = 0$, corresponding to the increasing and decreasing branches of the hysteresis cycle.}
\label{fig:distrHisteresis}
\end{figure}

The microscopic differences between the two dynamically stable stationary states are illustrated in Fig.~\ref{fig:distrHisteresis}, which compares the stationary wealth distributions obtained for $\gamma = 1.75$ and $\alpha = 0$, when the system is approached from below or from above in the hysteresis cycle shown in Fig.~\ref{fig:histGini}.

Although both distributions correspond to the same values of the control parameters, their structures are markedly different. Along the increasing-$\gamma$ branch, the distribution is more dispersed: a significant fraction of agents accumulates very low wealth, while a small number of agents reach substantially higher wealth values. In contrast, along the decreasing-$\gamma$ branch, the distribution is more concentrated around a typical wealth scale, with a few poorer agents and no pronounced accumulation at very high wealth. These structural differences are consistent with the separation between branches observed in the hysteresis curves, even though both cases correspond to strongly unequal states and reflect the existence of two dynamically stable stationary configurations selected by the system’s history.

\section{\B{Saddle-node theory of the upper spinodal}}
\label{app:saddlenode}

Here we derive the finite-$N$ upper spinodal $\gamma_c^{\rm up}(N)$ used in Sec.~\ref{subsec:spinodals} and Fig.~\ref{fig:spinodals}. The upper spinodal $\gamma_c^{\rm up}(N)$ is defined as the largest $\gamma$ for which a condensed configuration is dynamically self-sustained. We derive it as the saddle-node of a balance equation for the condensate share $\beta$: the condensate is stationary when its renormalization boost equals its tax drain, and the boundary of the bistable region is where this balance ceases to have a solution. The
construction has no free parameters: it follows from the same mean-field drift \eqref{eq:drift_mf} and self-consistent ergodic solution
\eqref{eq:rho_general}--\eqref{eq:Malpha} that describe the bulk, applied now to a configuration that also contains a condensate. 

\paragraph{Condensed configuration.}
In the condensed phase one or a few wealthy agents (the never-poor) hold a finite fraction $\beta$ of the total wealth, while the remaining agents form an ergodic bulk. Since the bulk shares only the fraction $1-\beta$ of the total wealth, and the replacement value is set by the global mean $\langle w\rangle$, the bulk experiences an effective redistribution strength
\begin{equation}
    \gamma_{\rm eff}=\frac{\gamma}{1-\beta}\;>\;\gamma .
    \label{eq:geff}
\end{equation}
Its stationary statistics are therefore those of the ergodic solution of
Sec.~\ref{subsec:meanfield} evaluated at $\gamma_{\rm eff}$: the boost
$\mu(\gamma_{\rm eff})$ and the moment $M_\alpha(\gamma_{\rm eff})$ are
obtained from Eqs.~\eqref{eq:norm}--\eqref{eq:Malpha} with $\gamma\to
\gamma_{\rm eff}$.

\paragraph{Condensate balance.}
In rescaled units the condensate carries wealth $u_c=\beta N$. The total $\alpha$-moment that enters the tax denominator, $M_\alpha^{\rm tot}=N^{-1}\sum_j u_j^\alpha$, receives a contribution from the bulk and one from the condensate,
\begin{equation}
    \bar{M}
    =(1-\beta)^{\alpha}\,M_\alpha(\gamma_{\rm eff})
     +\frac{(\beta N)^{\alpha}}{N},
    \label{eq:Mbar}
\end{equation}
where the prefactor $(1-\beta)^\alpha$ accounts for the rescaling of the bulk wealths by the share $1-\beta$.
Applying the drift \eqref{eq:drift_mf} to the condensate, $du_c/d\tau=(1-\beta)\,\mu(\gamma_{\rm eff})\,u_c-\gamma\,u_c^{\alpha}/(2\bar{M})$, the condensate is stationary when its renormalization boost balances its tax drain, $(1-\beta)\,\mu(\gamma_{\rm eff})\,u_c=\gamma\,u_c^{\alpha}/(2\bar{M})$, i.e.

\begin{equation}
    (1-\beta)\,\mu(\gamma_{\rm eff})
    =\frac{\gamma\,(\beta N)^{\alpha-1}}{2\,\bar{M}} .
    \label{eq:balance}
\end{equation}
The boost driving the condensate is the global expected minimum $\mathbb{E}[u_{\min}]=(1-\beta)\,\mu(\gamma_{\rm eff})$: the poorest agent is redrawn relative to the global mean, but it rejoins a bulk that holds only the share $1-\beta$, which rescales the bulk boost $\mu(\gamma_{\rm eff})$ accordingly.

Equation~\eqref{eq:balance}, together with \eqref{eq:geff} and \eqref{eq:Mbar}, implicitly determines the condensate share $\beta(\gamma)$. The left-hand side is the boost available from the bulk; the right-hand side is the (regressive) per-unit tax paid by the condensate, which for $\alpha<1$ decreases with the condensate wealth as $(\beta N)^{\alpha-1}$. This is the microscopic origin of condensation: a sufficiently rich agent pays a vanishing tax per unit wealth and cannot be brought back to the bulk.

\paragraph{Saddle-node and the spinodal.}
We write the residual of the balance \eqref{eq:balance} as $\mathcal{B}\,(\beta,\gamma)\equiv(1-\beta)\mu(\gamma_{\rm eff})-\gamma(\beta N)^{\alpha-1}/(2\bar{M})$, so that the stationary condensate shares are its roots $\mathcal{B}=0$. As $\beta\to0$ the per-unit tax diverges and $\mathcal{B}\to-\infty$. We find numerically that for fixed $\gamma$ below a threshold, $\mathcal{B}(\beta)$ rises to a positive maximum and falls again, crossing zero at two shares. At the lower one, $\mathcal{B}$ goes from negative to positive: it is an unstable ignition threshold, so a condensate below this share evaporates back into the bulk while a larger one grows. At the upper one, $\mathcal{B}$ goes from positive to negative: it is the stable, realized condensate, to which nearby shares relax. As $\gamma$ increases the two shares approach each other and merge when the maximum of $\mathcal{B}$ touches zero; for larger $\gamma$, $\mathcal{B}<0$ for every $\beta$ and no condensate can persist. This merging of a stable and an unstable root is a saddle-node, and the upper spinodal is the $\gamma$ at which it occurs
\begin{equation}
    \mathcal{B}(\beta,\gamma)=0,
    \qquad
    \frac{\partial \mathcal{B}}{\partial\beta}(\beta,\gamma)=0,
    \label{eq:saddlenode}
\end{equation}
which gives $\gamma_c^{\rm up}(N)$ and the share $\beta_{\rm sn}$ at which the condensate disappears. In practice we evaluate $\mu(\gamma_{\rm eff})$ and $M_\alpha(\gamma_{\rm eff})$ on a grid of $\gamma_{\rm eff}$ by solving Eqs.~\eqref{eq:norm}--\eqref{eq:Malpha}, and scan $\gamma$ upward until the roots of $\mathcal{B}$ disappear; the largest $\gamma$ with a root is $\gamma_c^{\rm up}(N)$. The procedure contains no adjustable parameters.

\begin{figure}[ht]
\centering
\includegraphics[width=\figwidth]{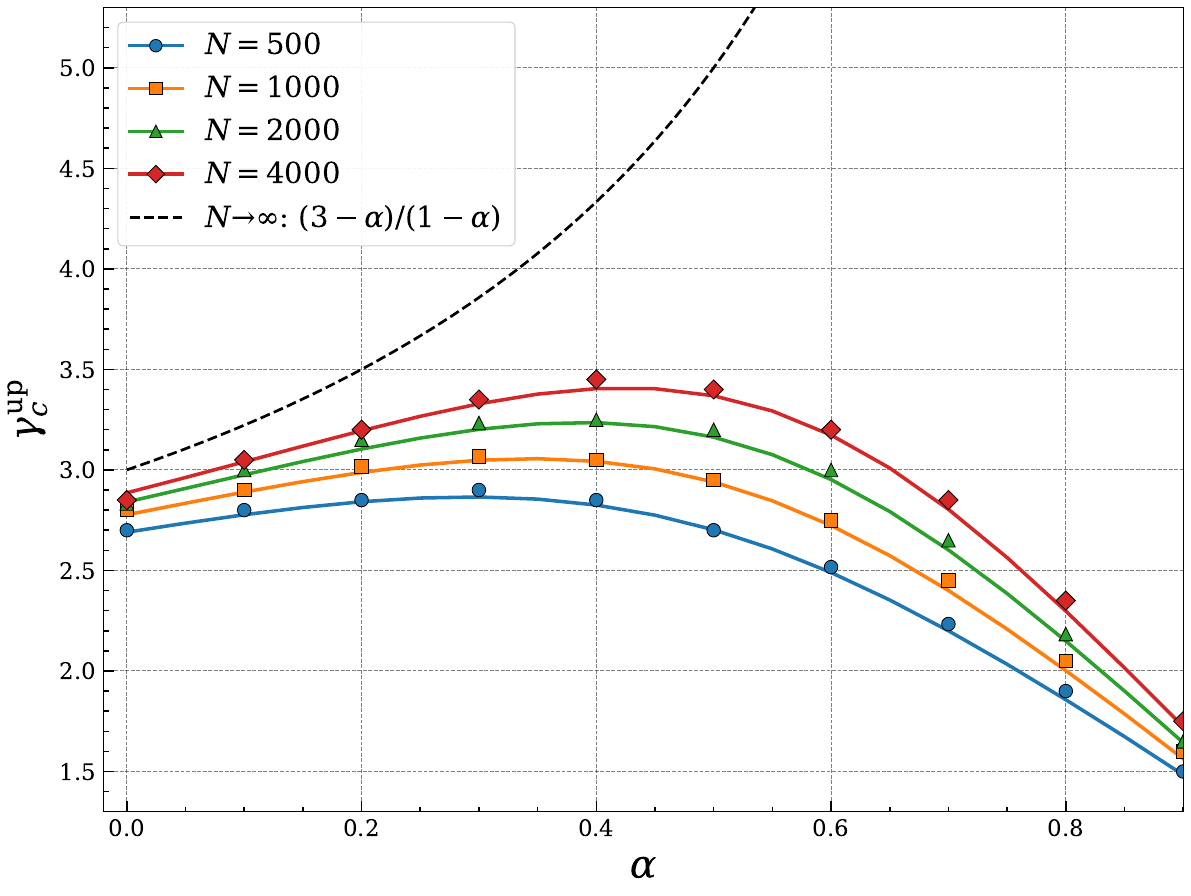}
\caption{%
    Upper spinodal $\gamma_c^{\rm up}(\alpha)$ for system sizes $N=\{500,1000,2000,4000\}$. Symbols: values measured with the adiabatic hysteresis protocol; solid lines: the parameter-free saddle-node theory [Eq.~\eqref{eq:saddlenode}]; black dashed: the thermodynamic limit $\gamma_{c,\infty}^{\rm up}=(3-\alpha)/(1-\alpha)$ [Eq.~\eqref{eq:upper_spinodal}].
}
\label{fig:upperN}
\end{figure}

\paragraph{Large-$N$ limit.}
Two competing requirements govern how far the high-inequality state can persist. To be taxed lightly, the condensate must be wealthy: its per-unit tax $(\beta N)^{\alpha-1}$ is regressive and falls as $u_c=\beta N$ grows. To leave the bulk unperturbed, and thus keep $\gamma_{\rm eff}=\gamma/(1-\beta)$ near $\gamma$, it must hold a small share $\beta$. At finite $N$ these pull against each other, since a small share $\beta$ implies a modest wealth $\beta N$; the saddle-node is their compromise and yields a spinodal below $\gamma_{c,\infty}^{\rm up}$. As $N\to\infty$ the tension dissolves, because a vanishing share can still carry a diverging wealth: we find $\beta_{\rm sn}\to0$ while $u_c=\beta_{\rm sn}N\to\infty$. Both simplifications then occur at once: $\gamma_{\rm eff}\to\gamma$ and the per-unit tax $(\beta N)^{\alpha-1}\to0$. The balance then reduces to $\mu(\gamma)=0$, i.e.\ the condition \eqref{eq:upper_spinodal_cond} of the main text, recovering
\begin{equation}
    \gamma_{c,\infty}^{\rm up}(\alpha)=\frac{3-\alpha}{1-\alpha}.
\end{equation}
This limit is approached slowly: the spinodal remains well below
$\gamma_{c,\infty}^{\rm up}$ even at the largest sizes considered
(Fig.~\ref{fig:upperN}). At finite $N$ the factor $(\beta N)^{\alpha-1}$ is positive and grows with $\alpha$, so Eq.~\eqref{eq:balance} is satisfied at a boost $\mu>0$, hence at $\gamma_{\rm eff}<\gamma_{c,\infty}^{\rm up}$ and $\gamma_c^{\rm up}(N)<\gamma_{c,\infty}^{\rm up}$. This downward correction increases
with $\alpha$, producing the non-monotonic $\gamma_c^{\rm up}(\alpha)$ of Fig.~\ref{fig:spinodals}. The resulting curve reproduces the upper
spinodal measured with the hysteresis protocol to within a few percent
over the whole range of $\alpha$, and the predicted size dependence $\gamma_c^{\rm up}(N)$ is confirmed by simulations at $N=500$--$4000$, as shown in Fig.~\ref{fig:upperN}.

We emphasize that this size dependence of the spinodal locations does not conflict with the size-independence reported in Appendix~\ref{app:finitesize}. The latter refers to the transition reached from the fixed uniform initial condition: this protocol selects a single, well-defined state inside the bistable region, and the corresponding transition point shows no systematic drift with $N$. The spinodals, by contrast, are the limits of metastability of the two branches, obtained by
adiabatically continuing each phase up to the point where it loses stability. These are distinct quantities: the fixed-initial-condition transition can remain $N$-independent while the spinodal boundaries that enclose it shift with $N$, widening the bistable window toward the finite interval $[\gamma_c^{\rm lo},\,(3-\alpha)/(1-\alpha)]$ in the thermodynamic limit.

\end{document}